\documentclass[12pt]{iopart}

\expandafter\let\csname equation*\endcsname\relax
\expandafter\let\csname endequation*\endcsname\relax
\usepackage{amsmath} 
\usepackage{amssymb} 
\usepackage{hyperref}
\usepackage{graphicx} 
\usepackage{mathrsfs} 
\usepackage{fullpage}
\usepackage[usenames,dvipsnames]{color} 
\usepackage{ulem}

\usepackage{caption}
\usepackage{subcaption}

\newcommand{\scri}{\mathscr{I}}

\usepackage[top=1in, bottom=1in, left=1.5in, right=1in]{geometry}
\numberwithin{equation}{section}

\begin{document}

\title{Gauge Invariant Spectral Cauchy Characteristic Extraction}
\author{Casey J. Handmer${}^{1}$, B\'{e}la Szil\'{a}gyi${}^{1}$,
  Jeffrey Winicour${}^2$} \address{${}^{1}$Theoretical Astrophysics
  350-17, California Institute of Technology, Pasadena, California
  91125, USA\\ ${}^{2}$Department of Physics and Astronomy University
  of Pittsburgh, Pittsburgh, PA 15260, USA}
\ead{chandmer@caltech.edu}

\begin{abstract}
We present gauge invariant spectral Cauchy characteristic
extraction. We compare gravitational waveforms extracted from a
head-on black hole merger simulated in two different gauges by two
different codes. We show rapid convergence, demonstrating both gauge
invariance of the extraction algorithm and consistency between the
legacy Pitt null code and the much faster Spectral Einstein Code
(\texttt{SpEC}).
\end{abstract}

\maketitle

\section{What is CCE? What is gravitational waveform gauge invariance?}

\begin{figure}[h!]
  \centering
  \includegraphics[width=0.5\textwidth]{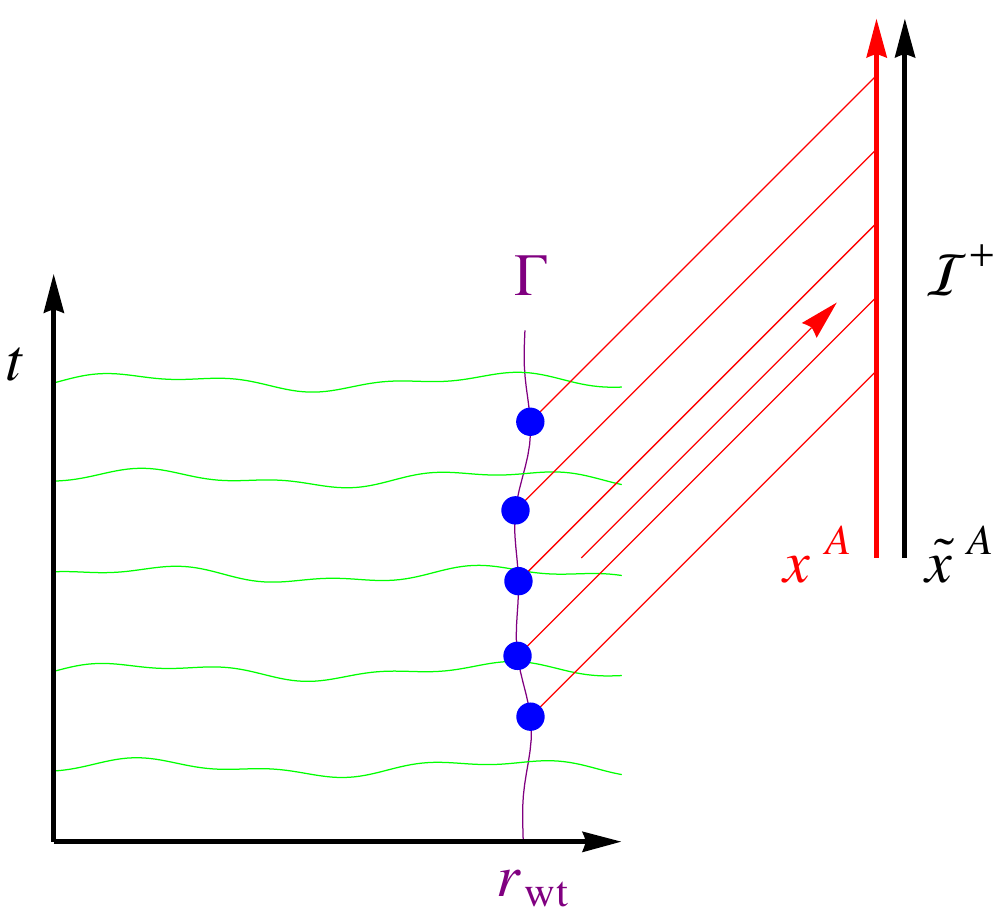}
  \caption{\small{Cauchy characteristic extraction. A Cauchy evolution
      of the Einstein field equation proceeds on a space-like
      foliation (green). A finite spheroidal worldtube \(\Gamma\) at
      areal radius \(r_{wt}\) forms the inner boundary to a
      characteristic evolution on a null foliation (red). Based on a
      coordinate system \(x^A\) derived from the Cauchy data on the
      worldtube, gravitational information is propagated to
      compactified future null infinity \(\scri^+\). At \(\scri^+\)
      an inertial coordinate system \(x^{\tilde A}\) is co-evolved, in
      which the desired gauge-invariant waveform can be expressed.}}
  \label{fig:AGNewsInertialDiagram}
\end{figure}

The strong gravitational radiation produced in the inspiral and merger
of binary black holes has been a dominant motivation for the
construction of gravitational wave observatories. The details of the
gravitational waveform supplied by numerical simulation is a key
theoretical tool to fully complement the sensitivity of the LIGO, Virgo, GEO, and KAGRA observatories, by enhancing detection and providing useful
scientific interpretation of the gravitational
signal\cite{Waldman2011,Accadia:2009zz,Grote:2010zz,Somiya:2012}.
Characteristic evolution coupled to Cauchy evolution via
Cauchy-characteristic extraction (CCE) provides the most accurate
numerical computation of the Bondi news function, which determines
both the waveform and the radiated energy and momentum at null
infinity.

In CCE, the Cauchy evolution is used to supply boundary data on a
timelike inner worldtube necessary to carry out a characteristic
evolution extending to future null infinity \(\scri^+\), where the
radiation is computed using the geometric methods developed by Bondi
{\it et al.}\cite{Bondi1962}, Sachs\cite{Sachs1962}, and
Penrose\cite{Penrose1963}, as depicted in
Fig.~\ref{fig:AGNewsInertialDiagram}. More intuitive methods,
including intrinsically inertial compactified hyperboloidal
formulations\cite{Bardeen2011a,Zenginoglu2008,Moncrief2009} have not
yet found adoption in the evolution of binary black holes. CCE is an
initial-boundary value problem based upon a timelike
worldtube\cite{TamburinoWinicour1966}. It has been implemented as a
characteristic evolution code, the Pitt null
code\cite{Isaacson:1983,Bishop:1997ik}, which incorporates a Penrose
compactification of the spacetime, and which has subsequently been
extended to higher order methods by Reisswig {\it et
  al.}\cite{Reisswig:2012ka}. It has more recently been implemented as
a spectral code within the Spectral Einstein Code (\texttt{SpEC}) by
Handmer and Szil\'{a}gyi\cite{Handmer:2014}, upon which the present
work is based.

One technical complication introduced by CCE is that the coordinates
induced on~\(\scri^+\) by the computational Cauchy coordinates on the
inner worldtube do not
correspond to inertial observers, i.e., to the
coordinates intrinsic to a distant freely falling and non-rotating
observatory. The gravitational waveform first obtained in the
``computational coordinates'' of CCE is in a scrambled form. This
gauge ambiguity in the waveform is
removed by constructing the transformation
between computational coordinates and inertial coordinates at \(\scri^+\).
There still remains the freedom in the choice of inertial observers.
In special relativistic theories, this freedom is reduced
to the translations and Lorentz transformations of the
Poincar{\' e} group. As explained in Sec.~\ref{sec:waveforms},
in an asymptotically flat space time the corresponding asymptotic symmetry
group consists of supertranslations and Lorentz transformations.
This freedom governs the redshift and initial phase of the waveform.

A physically relevant calculation of the radiation flux must also be
referred to such inertial coordinates at \(\scri^+\). In this paper,
the calculation of the energy-momentum flux via the Bondi news
function is first carried out in the induced worldtube coordinates
and then transformed to the inertial coordinates.

\section{Characteristic Formalism}
\label{sec:chform}

The characteristic formalism is based upon a family of outgoing null
cones emanating from an inner worldtube and extending to infinity
where they foliate \(\scri^+\) into spherical slices. We let \(u\) label
these hypersurfaces, \(y^A\) \((A=2,3)\) be angular coordinates that
label the null rays, and \(r\) be a surface area coordinate along the
outgoing null cones.

Employing the conventions used in \cite{Handmer:2014}, in the
resulting \(x^\alpha=(u,r,y^A)\) coordinates, the metric takes the
Bondi-Sachs form
\begin{align}
ds^2 = &-\left(e^{2\beta}(r W+1) - r^2 h_{AB} U^A U^B\right) du^2
\nonumber\\ &- 2 e^{2\beta} du dr - 2 r^2 h_{AB} U^B du dy^A + r^2
h_{AB} dy^A dy^B\;,
\label{eq:bmet}
\end{align}
where \(h^{AB}h_{BC}=\delta^A_C\) and \(\det(h_{AB})=\det(q_{AB})\), with
\(q_{AB}\) a unit sphere metric. In analyzing the Einstein equations,
we also use the intermediate variable
\begin{equation}
     Q_A = r^2 e^{-2\,\beta} h_{AB} U^B_{,r}\;.
\end{equation}

The metric coefficients \(W, h_{AB}, U^A, Q_A, \beta\) represent
respectively the mass aspect, the spherical 2-metric, the shift and
its radial derivative, and the lapse. The vector and tensor fields
\(h_{AB}, U^A, Q_A\) are expressed as spin-weighted fields by
contracting them with the complex dyad \(q^A\) for the unit sphere
metric satisfying \(q^Aq_A=0\), \(q^A\bar{q}_A = 2\), \(q^A =
q^{AB}q_B\), with \(q^{AB}q_{BC} = \delta^A_C\) and \(q_{AB} =
(q_A\bar{q}_B + \bar{q}_A q_B)/2\). Under this convention, the
spin-weighted functions \(U = U^A q_A\) and \(Q = Q_A q^A\), while \(J
= h_{AB}q^A q^B/2\) uniquely determines the spherical 2-metric component of the general 4-metric\cite{Bishop:1997ik}. We
chose a dyad consistent with the computational formulation of the
spin-weight raising \(\eth\) operator \cite{Gomez:1997}, given by
\(q^A=(-1,-i/\sin \theta)\) in standard spherical coordinates
\((\theta,\phi)\). This is regular everywhere except the poles, which
we can avoid through careful choice of grid points. It is worth noting
that any choice of angular coordinates are possible. Other conventions
use multiple patches to avoid singularities at the poles.

A key feature of the Bondi-Sachs formulation is that the Einstein
equations can be integrated along the outgoing characteristics in a
sequential order. We use a form which first
appeared in~\cite{Winicour1983} and was implemented as the Pitt code in~\cite{Bishop:1997ik,Bishop1998}:
\begin{align}
\beta,_r =&\; N_\beta\ \label{eq:beta},\\ 
(r^2 Q),_r =&\;-r^2(\bar{\eth} J + \eth K),_r + 2r^4 \eth(r^{-2}\beta),_r + N_Q\;,
\\ 
U,_r =&\; r^{-2} e^{2\beta} Q + N_U\;,\\ 
(r^2W),_r =&\; \frac{1}{2} e^{2\beta} \mathcal{R} -1 - e^\beta \eth \bar{\eth} e^\beta + \frac{1}{4} r^{-2} (r^4(\eth \bar{U} + \bar{\eth} U )),_r + N_W\;,\\ 
&\mathrm{and}\;\mathrm{the}\;\mathrm{evolution}\;\mathrm{equation} \nonumber \\
2(r J),_{ur} = & \left((1+r W )(rJ),_r\right),_r -r^{-1}(r^2 \eth U),_r + 2 r^{-1}e^\beta \eth^2 e^\beta - (r W),_r J +N_J\;,
\label{eq:jdot}
\end{align}
where
\begin{equation}
\mathcal{R} = 2K - \eth \bar{\eth} K + \frac{1}{2}(\bar{\eth}^2 J +
\eth^2 \bar{J}) + \frac{1}{4K} (\bar{\eth} \bar{J} \eth J -
\bar{\eth}J \eth \bar{J})\;,
    \label{eq:calR}
\end{equation}
is the curvature scalar associated with \(h_{AB}\), \(K^2 =1+ J \bar{J}\)
and \(N_\beta, N_Q, N_W,N_J\) are nonlinear terms given in
\cite{Bishop:1997ik}.

On each constant \(u\) hypersurface of the spacetime foliation, these
equations are integrated in turn. Given \(J\), \(\beta\) is solved,
then \(U\), \(Q\), and \(W\) in turn, enabling the computation of
\(J,_u\). \(J,_u\) permits a step forward in time and \(J\) is thus
defined on the next hypersurface. The radial compactification of
infinity is given by
\begin{equation}
   r = r_{wt} \rho/(1-\rho)\; ,\quad \frac{1}{2}\le \rho \le 1\;,
\label{eq:rcomp}
\end{equation}
where the compactification parameter \(r_{wt}(u,y^A)\) is the (not
necessarily constant) areal radius coordinate on the worldtube.

Angular derivatives are implemented using the action of the \(\eth\)
operator on spin-weighted spherical harmonics, e.g., \(\eth U = q^A q^B
U_{A:B}\), where a colon denotes the covariant derivative with respect
to \(q_{AB}\) \cite{Gomez:1997}. In spherical coordinates, this takes
the explicit form for a spin-weight-s field \(\eta\)
\begin{align}
\eth\eta = -(\sin^s \theta) \left(\frac{\partial}{\partial \theta} +
\frac{i}{\sin\theta}\frac{\partial}{\partial
  \phi}\right)(\sin^{-s}\theta\, \eta)\;,\\ \bar{\eth}\eta =
-(\sin^{-s} \theta) \left(\frac{\partial}{\partial \theta} -
\frac{i}{\sin\theta}\frac{\partial}{\partial
  \phi}\right)(\sin^s\theta\, \eta)\;.
\end{align}
 \(\bar{\eth}\) is the associated spin-weight lowering operator.

The spectral algorithm used to solve these equations and the treatment
of the nonlinear terms \(N_\beta, N_Q, N_U, N_W, N_J\) are detailed in
Handmer and Szil\'{a}gyi\cite{Handmer:2014}. Here, we extend the
characteristic spectral algorithm to calculating the gauge invariant
Bondi news at \(\scri^+\).

\section{Waveforms at \(\scri^+\)}
\label{sec:waveforms} 

For technical simplicity, the theoretical derivation of the waveform
at infinity is best presented in terms of an inverse surface-area
coordinate \(\ell=1/r\), where \(\ell=0\) at \(\scri^+\). In the resulting
\(x^\mu=(u,\ell,x^A)\) conformal Bondi coordinates, the physical
spacetime metric \(g_{\mu\nu}\) has the conformal compactification
\(\hat g_{\mu\nu}=\ell^{2} g_{\mu\nu}\), where \(\hat g_{\mu\nu}\) is
smooth at \(\scri^+\) and, referring to the metric (\ref{eq:bmet}),
takes the form\cite{TamburinoWinicour1966}
\begin{align}
   \hat g_{\mu\nu}dx^\mu dx^\nu=&\; -\left(e^{2\beta}(\ell^2 + \ell W)
   -h_{AB}U^AU^B\right)du^2 +2e^{2\beta}dud\ell \\
   &\;-2 h_{AB}U^Bdudx^A + h_{AB}dx^Adx^B\;. \nonumber
   \label{eq:lmet}
\end{align}

As described in~\cite{BabiucEtAl2008,Babiuc:2010ze}, both the Bondi news
function \(N(u,x^A)\) and the Newman-Penrose Weyl tensor
component~\cite{Newman1962}
\begin{equation}
\Psi_4^0(u,x^A)=\lim_{r\rightarrow \infty} r \psi_4\;,
\end{equation}
which describe the waveform, are determined by the asymptotic
limit at \(\scri^+\) of the tensor field
\begin{equation}
 \hat \Sigma_{\mu\nu} = \frac{1}{\ell}(\hat \nabla_\mu\hat \nabla_\nu
 -\frac{1}{4}\hat g_{\mu\nu} \hat \nabla^\alpha\hat
 \nabla_\alpha)\ell\;.
\label{eq:Sigma}
\end{equation}
This limit is constructed from the leading coefficients in an
expansion of the metric about~\(\scri^+\) in powers of \(\ell\). We thus
write
\begin{equation}
   h_{AB}= H_{AB}+\ell c_{AB}+O(\ell^2)\;.
\end{equation}
Conditions on the asymptotic expansion of the remaining components of
the metric follow from the Einstein equations:
\begin{equation}
    \beta=H+ O(\ell^2)\;,
\end{equation}
\begin{equation}
    U^A= L^A+2\ell e^{2H} H^{AB}D_B H+O(\ell^2)\;,
\end{equation}
and
\begin{equation}
    W = D_A L^A +\ell (e^{2H}{\cal R}/2 +D_A D^A e^{2H} -
    1)+O(\ell^2)\;,
\end{equation}
where \(H\) and \(L\) are the asymptotic limits of \(\beta\) and \(U\) and
where \({\cal R}\) and \(D_A\) are the 2-dimensional curvature scalar and
covariant derivative associated with \(H_{AB}\).

The expansion coefficients \(H\), \(H_{AB}\), \(c_{AB}\), and \(L^A\) (all
functions of \(u\) and \(x^A\)) completely determine the radiation
field. One can further specialize the Bondi coordinates to be {\em
  inertial} at \(\scri^+\), i.e., have Minkowski form, in which case
\(H=L^A=0\), \(H_{AB}=q_{AB}\) (the unit sphere metric) so that the
radiation field is completely determined by \(c_{AB}\). However, the
characteristic extraction of the waveform is carried out in
computational coordinates (determined by the Cauchy data on the
extraction worldtube) so this inertial simplification cannot be
assumed.

In order to first compute the Bondi news function in the \(\hat
g_{\mu\nu}\) computational frame, it is necessary to determine the
conformal factor \(\omega\) relating \(H_{AB}\) to a unit sphere metric
\(Q_{AB}\), i.e., to an inertial conformal Bondi
frame\cite{TamburinoWinicour1966} satisfying
\begin{equation}
          Q_{AB}=\omega^2H_{AB}\;.
\label{eq:unsph}
\end{equation}
(See~\cite{Winicour1987} for a discussion of how the news in an
arbitrary conformal frame is related to its expression in this
inertial Bondi frame.) We can determine \(\omega\) by solving the
elliptic equation governing the conformal transformation of the
curvature scalar (\ref{eq:calR}) to a unit sphere geometry:
\begin{equation}
     {\cal R}=2(\omega^2+H^{AB}D_A D_B \log \omega)\;.
\label{eq:conf}
\end{equation}
The elliptic equation (\ref{eq:conf}) need only be solved at the
initial time where, with initial data \(J|_{\scri^+}=0\), \(H^{AB}D_A
D_B\) simplifies to the 2-Laplacian on the unit sphere. Then, as
described in the next section, application of the Einstein equations
on \(\scri^+\) determines the time dependence of \(\omega\) according to
\begin{equation}
     2\hat n^\alpha \partial_{\alpha} \log \omega =-e^{-2H}D_AL^A\;,
\label{eq:omegadot}
\end{equation}
where \(\hat n^\alpha =\hat g^{\alpha\beta}\nabla_\beta \ell\) is the
null vector tangent to the generators of \(\scri^+\). We use
(\ref{eq:omegadot}) to evolve \(\omega\) along the generators of
\(\scri^+\) given a solution of (\ref{eq:conf}) as initial condition.

First recall some basic elements of Penrose compactification. In a
general conformal frame with metric \( \hat g_{\mu\nu}=\Omega^2
g_{\mu\nu}\), where \(\Omega=0\) on \(\scri^+\), the vacuum Einstein
equations \(G_{\mu\nu}=0\) take the form
\begin{equation}
   \Omega^2 \hat G_{\mu\nu} + 2\Omega \hat \nabla_\mu \hat \nabla _\nu
   \Omega - \hat g_{\mu\nu} \bigg (2\Omega \hat \nabla^\rho \hat
   \nabla_\rho \Omega -3( \hat \nabla^\rho \Omega \hat \nabla_\rho
   \Omega) \bigg ) =0\;.
\end{equation}
It immediately follows that
\begin{equation}
   (\hat \nabla^\rho \Omega ) \hat \nabla_\rho \Omega|_{\scri^+} = 0\;,
\end{equation}  
 so that \(\scri^+\) is a null hypersurface and that
\begin{equation}
     [ \hat \nabla_\mu \hat \nabla _\nu \Omega -\frac{1}{4}\hat
       g_{\mu\nu} \hat \nabla^\rho \hat \nabla_\rho \Omega
     ]|_{\scri^+} =0\;.
\end{equation} 

With respect to this frame, the construction of an inertial conformal
frame proceeds as follows. We introduce a new conformal factor \(\tilde
\Omega =\omega \Omega\), with \(\tilde g_{\mu\nu} =\omega^2 \hat
g_{\mu\nu}\) by requiring, in accord with (\ref{eq:omegadot}),
\begin{equation}
  [2 \hat n^\sigma \partial_\sigma \omega +\omega \hat \nabla_\sigma
    \hat n^\sigma]|_{\scri^+} =0\;, \quad \hat n^\sigma =\hat
  g^{\rho\sigma}\nabla_\rho \Omega\;.
      \label{eq:omev}
\end{equation}
As a result, it follows from a straightforward calculation that
\begin{equation}
    \tilde \nabla^\rho \tilde \nabla_\rho \tilde \Omega |_{\scri^+}
    = 0\;,
    \label{eq:divfr}
\end{equation}
i.e., in the \(\tilde g_{\mu\nu}\) conformal frame \(\scri^+\) is null,
shear- and divergence-free.

It also follows that
\begin{equation}
      \tilde n^\sigma \tilde \nabla _\sigma \tilde n^\nu |_{\scri^+}
      =0\;,
\end{equation} 
where \(\tilde n^\sigma =\tilde g^{\rho\sigma}\tilde\nabla_\rho
\tilde\Omega\), i.e., in the \(\tilde g_{\mu\nu}\) frame, \(\tilde
n^\sigma\) is an affinely parametrized null generator of~\(\scri^+\).

To construct inertial coordinates \((\tilde u, x^{\tilde A})\) on
\(\scri^+\), we first assign angular coordinates \(x^{\tilde A}\) to each
point of the initial spacelike spherical slice \(u=u_0\) of \(\scri^+\).
We then propagate these coordinates along the generators of \(\scri^+\)
according to
\begin{equation}
      \tilde n^\rho \partial_\rho x^{\tilde A} |_{\scri^+} = \omega^{-1}
      \hat n^\rho \partial_\rho x^{\tilde A} |_{\scri^+} = 0\;.
\end{equation}
In addition, we require
\begin{equation}
      \tilde n^\rho \partial_\rho \tilde u |_{\scri^+} = \omega^{-1}
      \hat n^\rho \partial_\rho \tilde u |_{\scri^+} = 1\;,
\end{equation}
so that \(\tilde u\) is an affine parameter along the generators in the
\(\tilde g_{\mu\nu}\) conformal frame.

\section{News}
The Bondi news function \(N\) is computed in the computational
coordinates with the appropriate conformal transformation. It is then
interpolated onto the inertial coordinates. The formalism follows that
of \cite{Bishop:1997ik}, Appendix~B (with a sign error in \(s_3\)
corrected):
\begin{equation}
N = \frac{1}{4 \omega A}\left( s_1 + s_2 + \frac{1}{4}\left(\eth
\bar{U} + \bar{\eth} U \right) s_3 - 4 \omega^{-2} s_4 + 2 \omega^{-1}
s_5 \right),
\end{equation}
where \(A = \omega e^{2 \beta}\) and the \(s_i\) terms are
\begin{align}
s_1& = (J^2 \bar J_{,\ell u} + J \bar J J_{,\ell u} - 2 J K K_{,\ell
  u} - 2 J K_{,\ell u} + 2 J_{,\ell u} K + 2 J_{,\ell u})/(K + 1)
\;,\nonumber \\ s_2& = (\eth J_{,\ell} J \bar J \bar U + 2 \eth
J_{,\ell} K \bar U + 2 \eth J_{,\ell} \bar U + \eth \bar J_{,\ell} J^2
\bar U \nonumber \\ & - 2 \eth K_{,\ell} J K \bar U - 2 \eth K_{,\ell}
J \bar U + 2 \eth U J \bar J K_{,\ell} - 2 \eth U J \bar J_{,\ell} K
\nonumber \\ & - 2 \eth U J \bar J_{,\ell} + 4 \eth U K K_{,\ell} + 4
\eth U K_{,\ell} + 2 \eth \bar U J \bar J J_{,\ell} - 2 \eth \bar U J
K K_{,\ell} \nonumber \\ & - 2 \eth \bar U J K_{,\ell} + 4 \eth \bar U
J_{,\ell} K + 4 \eth \bar U J_{,\ell} + \bar \eth J_{,\ell} J \bar J U
+ 2 \bar \eth J_{,\ell} K U + 2 \bar \eth J_{,\ell} U \nonumber \\ & +
\bar \eth \bar J_{,\ell} J^2 U - 2 \bar \eth K_{,\ell} J K U - 2 \bar
\eth K_{,\ell} J U + 2 \bar \eth U J^2 \bar J_{,\ell} - 2 \bar \eth U
J K K_{,\ell} \nonumber \\ & - 2 \bar \eth U J K_{,\ell} + 2 \bar \eth
\bar U J^2 K_{,\ell} - 2 \bar \eth \bar U J J_{,\ell} K - 2 \bar \eth
\bar U J J_{,\ell})/(2 (K + 1)) \;,\nonumber \\
s_3& =-(J^2 \bar J_{,\ell} + J \bar J J_{,\ell} - 2 J K K_{,\ell} - 2
J K_{,\ell} + 2 J_{,\ell} K + 2 J_{,\ell})/(K + 1) \;,\nonumber \\ s_4&
=(\eth A \eth \omega J \bar J + 2 \eth A \eth \omega K + 2 \eth A \eth
\omega - \eth A \bar \eth \omega J K \nonumber \\ & - \eth A \bar \eth
\omega J - \eth \omega \bar \eth A J K - \eth \omega \bar \eth A J +
\bar \eth A \bar \eth \omega J^2)/(2 (K + 1)) \;,\nonumber \\ s_5& =(2
\eth^2 A J \bar J + 4 \eth^2 A K + 4 \eth^2 A + 2 \bar \eth^2 A J^2 -
4 \bar \eth \eth A J K \nonumber \\ & - 4 \bar \eth \eth A J + \eth A
\eth J J \bar J^2 + 2 \eth A \eth J \bar J K + 2 \eth A \eth J \bar J
+ \eth A \eth \bar J J^2 \bar J \nonumber \\ & + 2 \eth A \eth \bar J
J K + 2 \eth A \eth \bar J J - 2 \eth A \eth K J \bar J K - 4 \eth A
\eth K J \bar J - 4 \eth A \eth K K \nonumber \\ & - 4 \eth A \eth K -
\eth A \bar \eth J J \bar J K + 2 \eth A \bar \eth J K + 2 \eth A \bar
\eth J - \eth A \bar \eth \bar J J^2 K \nonumber \\ & + 2 \eth A \bar
\eth K J^2 \bar J - \eth J \bar \eth A J \bar J K - 2 \eth J \bar \eth
A J \bar J - 2 \eth J \bar \eth A K \nonumber \\ & - 2 \eth J \bar
\eth A - \eth \bar J \bar \eth A J^2 K - 2 \eth \bar J \bar \eth A J^2
+ 2 \eth K \bar \eth A J^2 \bar J \nonumber \\ & + 4 \eth K \bar \eth
A J K + 4 \eth K \bar \eth A J + \bar \eth A \bar \eth J J^2 \bar J
\nonumber \\ & + \bar \eth A \bar \eth \bar J J^3 - 2 \bar \eth A \bar
\eth K J^2 K)/(4 (K + 1)) \;.
\end{align}

In our implementation, \(,_l\) derivatives are derived from spectrally
calculated \(,_\rho\) derivatives using the appropriate Jacobian.

\section{Results}

In our comparison tests of CCE, the worldtube boundary data were
extracted from a simulation of an equal mass non-spinning head-on
black hole collision, with initial separation of \(30M\). The control
run (Isotropic) utilized the standard harmonic gauge damping identical
to that in \cite{Szilagyi:2009qz} throughout the head-on merger and
ring-down. Harmonic gauge damping adds a dissipative forcing term to
the wave equations satisfied by the harmonic Cartesian spatial
coordinates \((x,y,z)\). In order to diminish the effects of a custom
designed gauge, we also compare with results of another run (HytZero)
which turns off gauge damping in the harmonic \(y\)-direction,
transverse to the \(x\)-direction motion of the black holes. These two
high-resolution runs were used as boundary data for all the subsequent
CCE runs. These runs include 3 different resolutions, 2 different codes, 2 different gauges, and 3 different extraction radii, for a total of 36 runs.

As described in \cite{Handmer:2014}, the \texttt{SpEC} characteristic
evolution algorithm exploits spectral methods and innovative integral
methods that greatly improve upon the speed and accuracy of the Pitt
null code. This is seen as essential for for taking advantage of the
efficiency of \texttt{SpEC} Cauchy evolution. The necessary
improvement in efficiency has been preserved in the \texttt{SpEC}
extraction module, as displayed in Table~\ref{tab:ALLTHETHINGS}. The
comparison runs were performed using the current version of the Pitt
code~\cite{Babiuc:2010ze}, which forms part of the Einstein Toolkit.

The initial conditions and extraction parameters were deliberately
chosen as a stressful test of the algorithms. In particular, at the
beginning of the run the black hole excision boundary extends out to
Cartesian radius \(R=16M\), which is very close to our smallest choice
of extraction radius at \(R=30M\). At this radius, gauge effects are
highly significant and would make perturbative extraction schemes
meaningless, in accordance with our intentions. One consequence of
such an extreme choice is that differences between the Pitt and
\texttt{SpEC} inertial frame and worldtube initialization procedures lead
to noticeably different waveforms. Worldtube initialization involves
supplying the ``integration constants'' from the Cauchy code, which
allows radial integration of the characteristic hypersurface and evolution
equations (\ref{eq:beta}) -- (\ref{eq:jdot}) from the worldtube to
\(\scri^+\). In both Pitt null code and SpEC, the initial condition on \(J\) is determined by the inner boundary value, supplied by the Cauchy evolution, with a smooth roll off to zero at \(\scri^+\). 

The extraction worldtube \(\Gamma\) is determined by a surface of
constant Cartesian radius~\(R\). In the Pitt CCE code, the areal radius
\(r_{wt}\) of \(\Gamma\) lies between two surfaces of constant Cartesian
radii \(R_1\le r_{wt} \le R_2\) and this carries over to the
compactified radial coordinate. As a result, interpolation is
necessary to supply the integration constants, which introduces
numerical error. In the \texttt{SpEC} CCE code, this interpolation
error is avoided by introducing the compactified radial coordinate
(\ref{eq:rcomp}), with range \(1/2\le\rho\le1\) between \(\Gamma\)
and \(\scri^+\).

\begin{figure}[h!]
    \centering
    \includegraphics[width=0.8\textwidth]{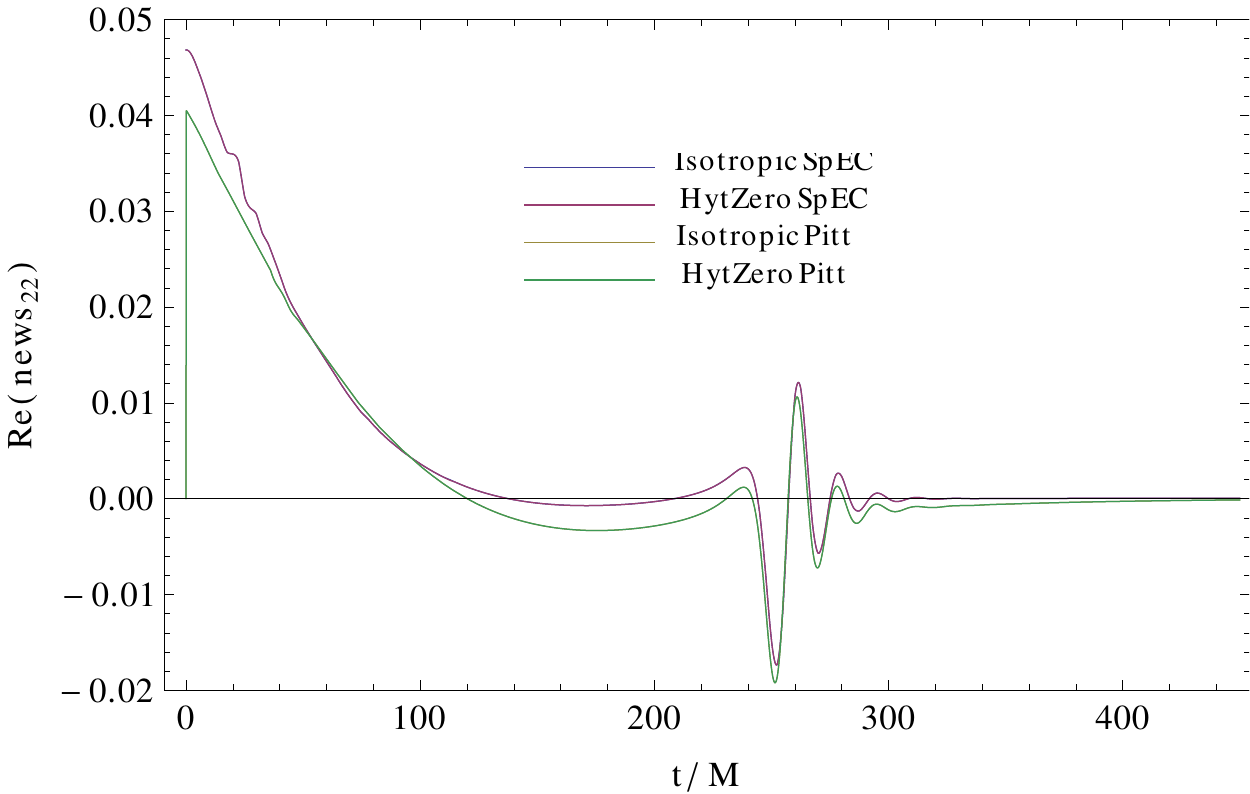}
    \caption{\small{Waveforms of the real part of the \((2,2)\) spherical
        harmonic mode of the news function, as computed by the
        Pitt null code and \texttt{SpEC} with extraction worldtube at
        \(R=30M\). Different initialization procedures at the worldtube
        give rise to a difference between the Pitt and \texttt{SpEC}
        waveforms, which is most pronounced at this small extraction
        radius. The different gauge choices, Isotropic and HytZero, do
        not have noticeable effect on this scale, indicating successful
        gauge effect removal in both codes.}}
    \label{fig:AGNewsWaveform30}
\end{figure}

\begin{figure}[h!]
    \centering
    \includegraphics[width=0.8\textwidth]{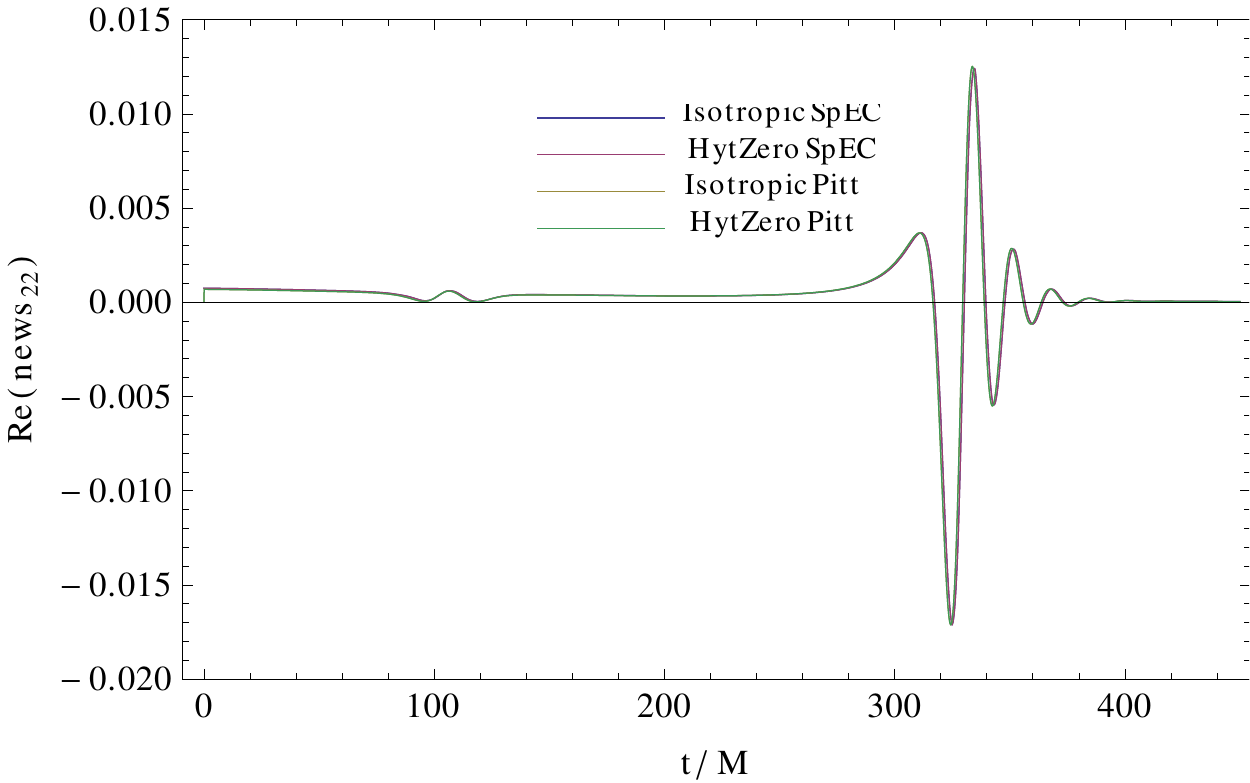}
    \caption{\small{Waveforms of the real part of the \((2,2)\) mode of the
        news function, as computed by the Pitt and
        \texttt{SpEC} codes with extraction worldtube at \(R=100M\). Compared
        to Fig.~\ref{fig:AGNewsWaveform30}, at this larger extraction
        radius the worldtube initialization differences lead to a much
        smaller difference between waveforms, which appear nearly
        identical here. The main discrepancy arises from the treatment
        of the junk radiation at early times. Here, too, gauge
        differences between HytZero and
        Isotropic are not visible at this scale.}}
    \label{fig:AGNewsWaveform100}
\end{figure}

\begin{figure}[h!]
  \centering
    \includegraphics[width=0.8\textwidth]{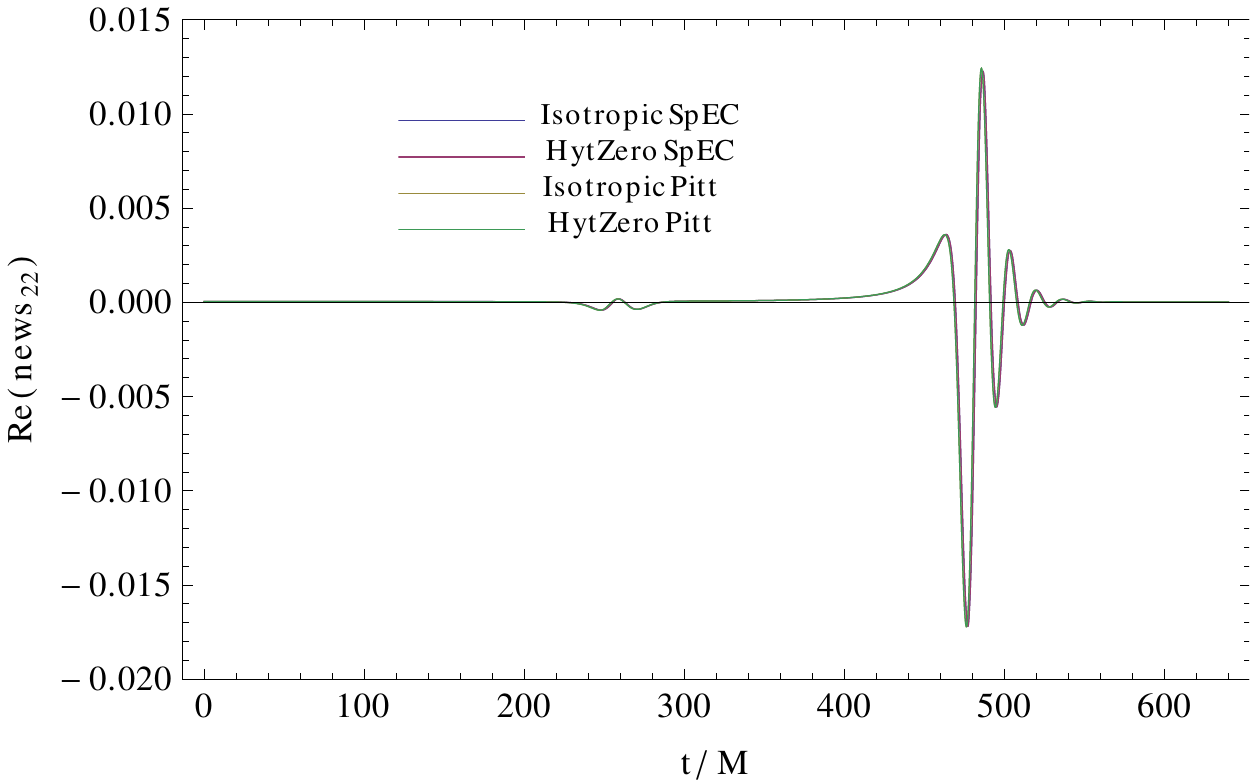}
    \caption{\small{Waveforms of the real part of the \((2,2)\) mode of the
        news function, as computed by the Pitt and
        \texttt{SpEC} codes with extraction worldtube at \(R=250M\). At this
        large extraction radius there is a barely noticeable difference
        between all the waveforms, limited to the junk radiation at
        early times.}}
    \label{fig:AGNewsWaveform250}
  \end{figure}

Worldtube data from each run were extracted using both Pitt and
\texttt{SpEC} CCE, at three different Cartesian radii: \(R=30M\),
\(R=100M \) and \(R=250M\), as illustrated by the news function
waveforms in Figs.~\ref{fig:AGNewsWaveform30},
\ref{fig:AGNewsWaveform100}, and \ref{fig:AGNewsWaveform250},
respectively. In these figures, the HytZero and Isotropic waveforms
are so close that they appear on top of one another. The major discrepancy
between the Pitt and \texttt{SpEC} waveforms is due to the worldtube
interpolation error in the Pitt code. This is especially evident at small extraction radii,
where there is strong ``junk'' radiation near the worldtube, which is
inherent in the initial Cauchy data and its mismatch with the initial
characteristic data.

\begin{figure}[h!]
    \centering
    \includegraphics[width=0.8\textwidth]{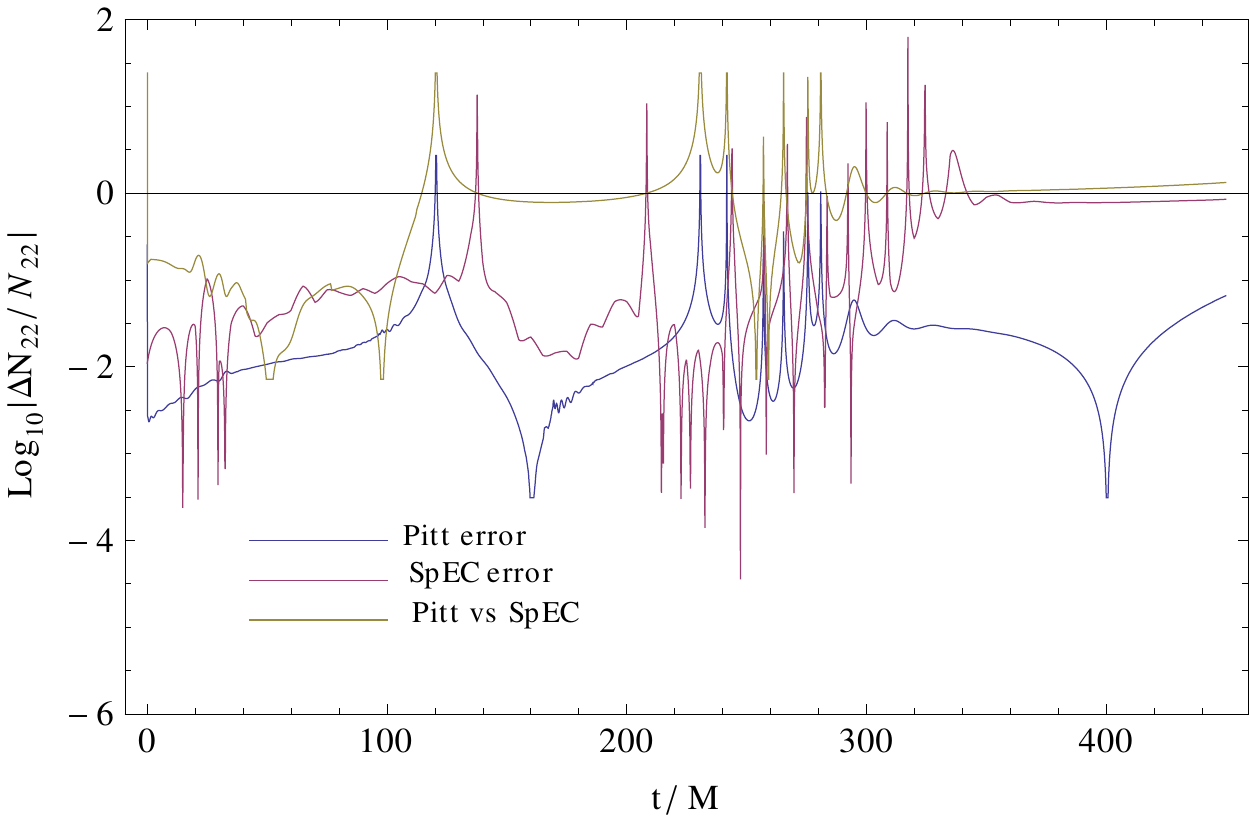}
    \caption{\small{Graphs showing the relative difference between the
        real part of the \((2,2)\) mode Pitt and \texttt{SpEC} news
        function waveforms for extraction radius \(R=30M\), in
        comparison to the relative numerical error implied by
        convergence tests, corresponding to the waveforms in
        Fig.~\ref{fig:AGNewsWaveform30}. While both \texttt{SpEC} and
        Pitt have comparable and consistent levels of error, the codes
        do not agree within that level of error at this extraction
        radius.}}
    \label{fig:AGPittvsSpECnews30}
\end{figure}

\begin{figure}[h!]
    \centering
    \includegraphics[width=0.8\textwidth]{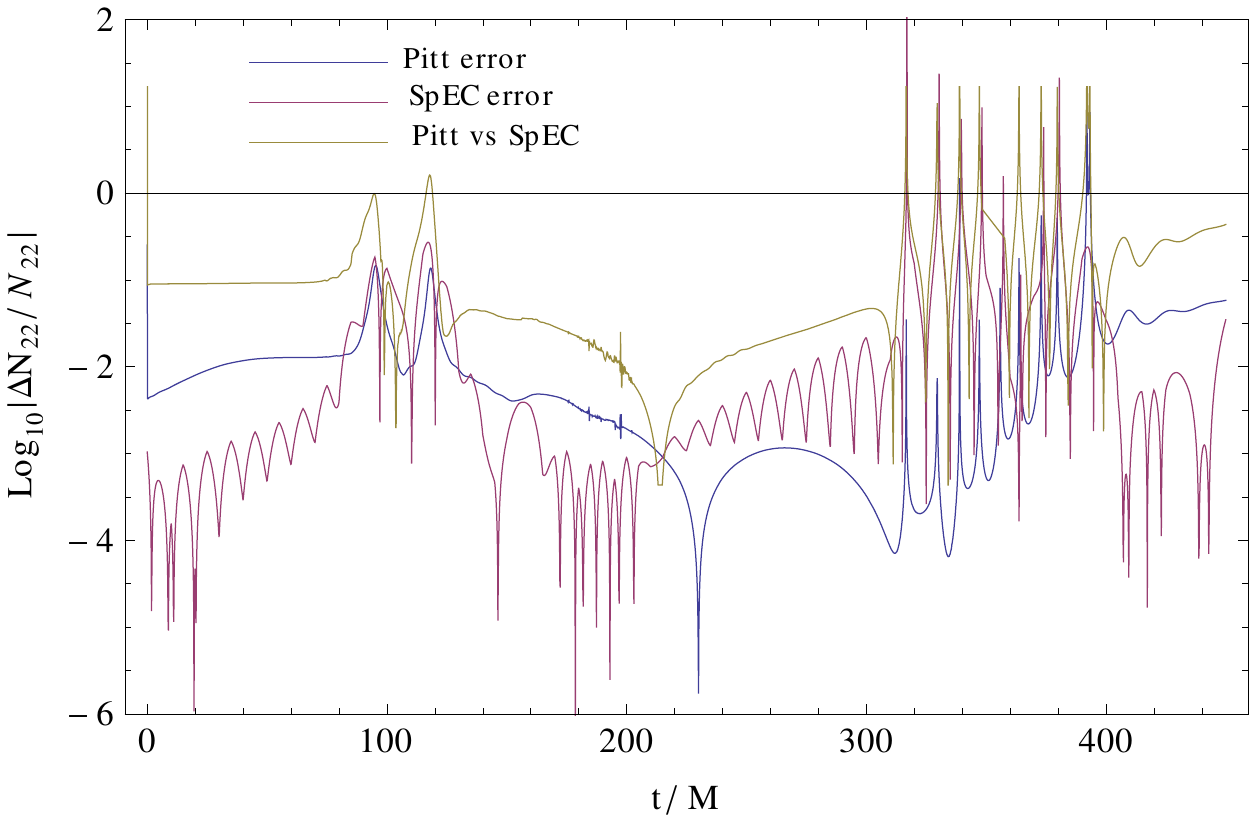}
    \caption{\small{Graph showing the relative difference between
        the real part of the \((2,2)\) mode \texttt{SpEC} and Pitt news
        waveforms for extraction radius \(R=100M\), and the relative
        numerical error, corresponding to the waveforms in
        Fig.~\ref{fig:AGNewsWaveform100}. In comparison with Fig.
        \ref{fig:AGPittvsSpECnews30}, by \(R=100M\) the difference
        between the Pitt and \texttt{SpEC} algorithms has dropped to the
        level of numerical error in each algorithm.}}
    \label{fig:AGPittvsSpECnews100}
\end{figure}

\begin{figure}[h!]
    \centering
    \includegraphics[width=0.8\textwidth]{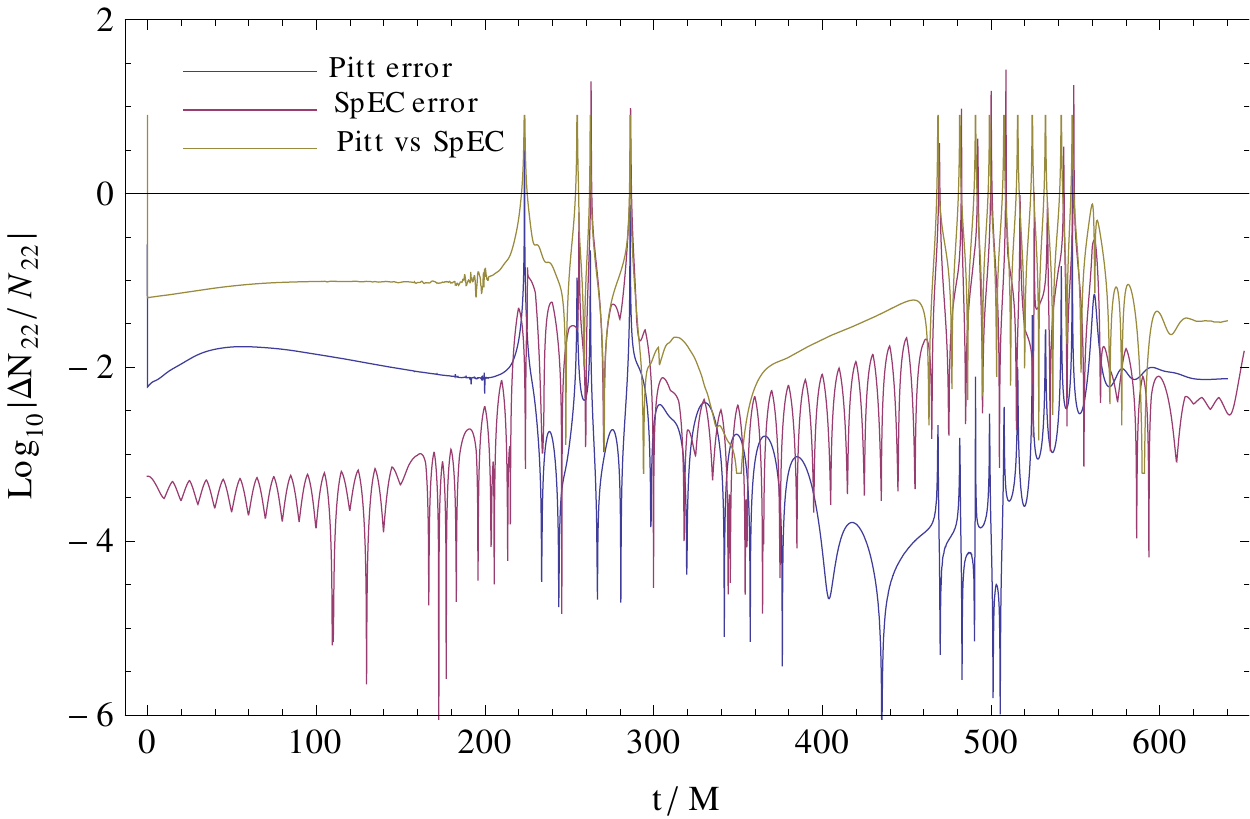}
    \caption{\small{Graphs showing the relative difference between \texttt{SpEC}
        and Pitt news waveforms at \(R=250M\), in comparison to the
        relative numerical error, corresponding to the waveforms in
        Fig.~\ref{fig:AGNewsWaveform250}. During the post-junk part of
        the waveform, the error due to worldtube initialization has
        dropped to the level of the numerical error, completing the trend seen in
        Fig.~\ref{fig:AGPittvsSpECnews100}.}}
    \label{fig:AGPittvsSpECnews250}
\end{figure}

This interpolation error in the Pitt code converges away at larger
radii, where the field gradients between \(R_1\) and \(R_2\) become
smaller. This is seen in Figs.~\ref{fig:AGPittvsSpECnews30},
\ref{fig:AGPittvsSpECnews100}, and \ref{fig:AGPittvsSpECnews250}, where
the relative difference between the Pitt and \texttt{SpEC} news function
waveforms is compared with the relative numerical error implied by convergence tests.

Each run was computed at 3 different resolutions to monitor
convergence, as indicated in Table~\ref{tab:ALLTHETHINGS}. In the
following subsections, we first show convergence and the removal of
gauge effects, separately for the Pitt and \texttt{SpEC} codes. Next,
we compare \(\Psi_4^0\) waveforms and establish further agreement
between the two codes. Finally, we examine the evolution of the
inertial coordinates at \(\scri^+\) relative to the worldtube
coordinates induced by the Cauchy evolution.

Comparison of the relative error \(E_{rel}\) between dataset \(A\) and
dataset \(B\) is computed according to
\begin{equation}
E_{rel} = \log_{10}\left(\frac{|A-B|}{|B|}\right) \;,
\end{equation}
where in convergence tests \(B\) is the highest resolution dataset,
and the real parts of the \((\ell,m)=(2,2)\) spherical harmonic modes
are compared.

\begin{table}
\centering
  \begin{tabular}{| r | c c c | c c c |}
    \hline Run & Pitt1 & Pitt2 &Pitt3 & \texttt{\texttt{SpEC}}1 &
    \texttt{\texttt{SpEC}}2 & \texttt{\texttt{SpEC}}3 \\ \hline
    \(N_r\) & 100 & 150 & 200 & 10 & 12 & 14 \\ \(N\) or \(L\) & 40 &
    60 & 80 & 12 & 14 & 17 \\ \(\Delta t/M\) & 0.1 & \(0.0666\dots\) &
    0.05 & 1.0 & \(0.666\dots\) & 0.5 \\ T (CPU hours) & 173 & 274 &
    374 & 0.7 & 1.9 & 3.1 \\ \hline
  \end{tabular}
  \caption{\small{Resolution parameters used for code convergence
      comparisons, with time steps \(\Delta t\). \(N_r\) represents the
      radial grid sizes. The Pitt null code uses two stereographic
      patches with \(2 N^2\) total number of angular grid points. The
      \texttt{SpEC} code has \(2 L^2\) total angular grid points. T
      is the CPU time taken for \(R=30M\), \(t_{final}=450M\) runs
      in the Isotropic gauge, and is representative for the other runs. All resolutions and codes were run from the same initial data.}}
  \label{tab:ALLTHETHINGS}
\end{table}



\subsection{Pitt code convergence and removal of gauge effects}
\label{sec:pitt}

\begin{figure}[h!]
  \centering
  \includegraphics[width=0.8\textwidth]{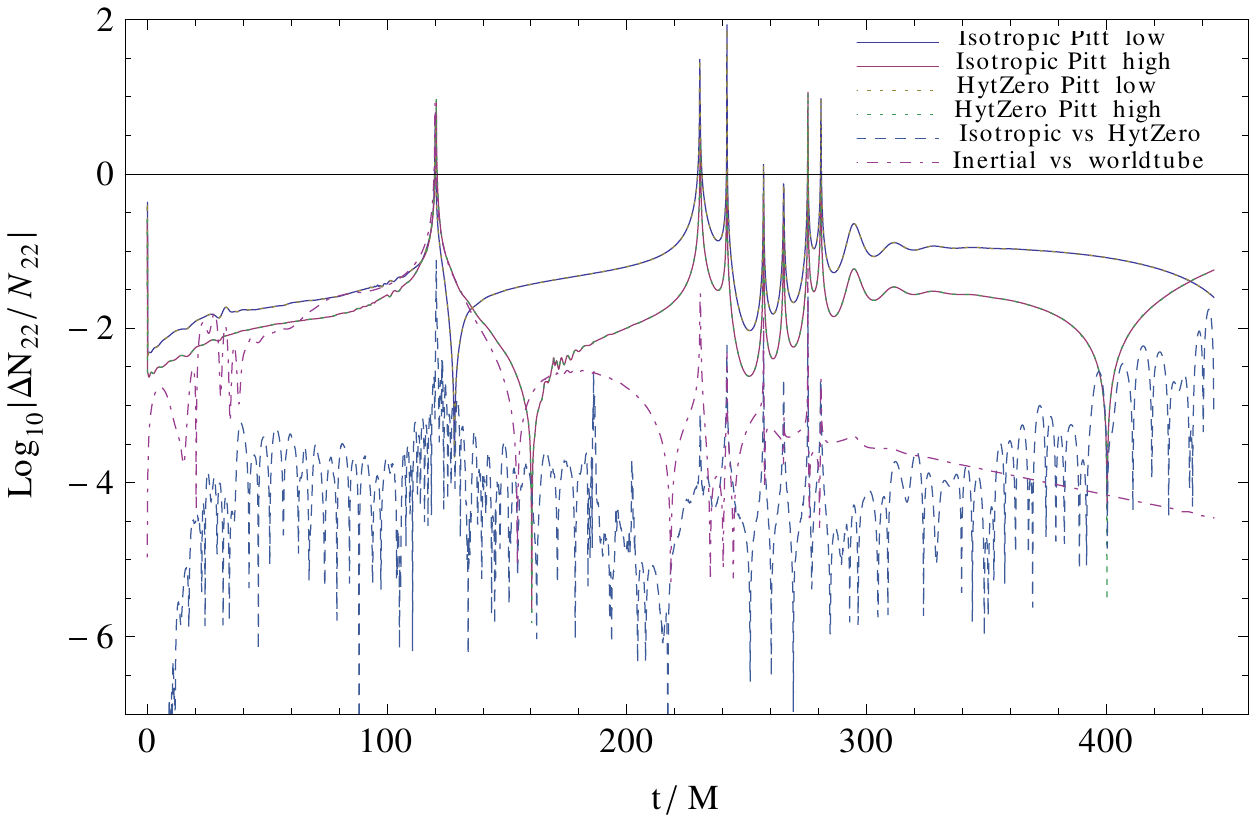}
  \caption{\small{Graphs of the relative error \(\log_{10}|\Delta
      N_{22}/N_{22}|\) in the \((2,2)\) mode of the news function for the
      \(R=30M\) Pitt extraction run in both gauges. The relative
      errors for the Pitt1 (low) and Pitt2 (high) resolutions
      (compared to Pitt2 and Pitt3 respectively) are rescaled to
      demonstrate convergence. The dashed blue line indicates the
      relative error (Isotropic vs HytZero) between the news computed
      in both gauges. The dot-dashed purple line (Inertial vs
      worldtube) indicates the relative error between the news
      computed in the worldtube coordinates and the inertial
      coordinates. At this small extraction radius, this discrepancy
      is high due to the strong gauge effect of junk radiation.}}
    \label{fig:AGPittConvNews30}
\end{figure}

\begin{figure}[h!]
  \centering
  \includegraphics[width=0.8\textwidth]{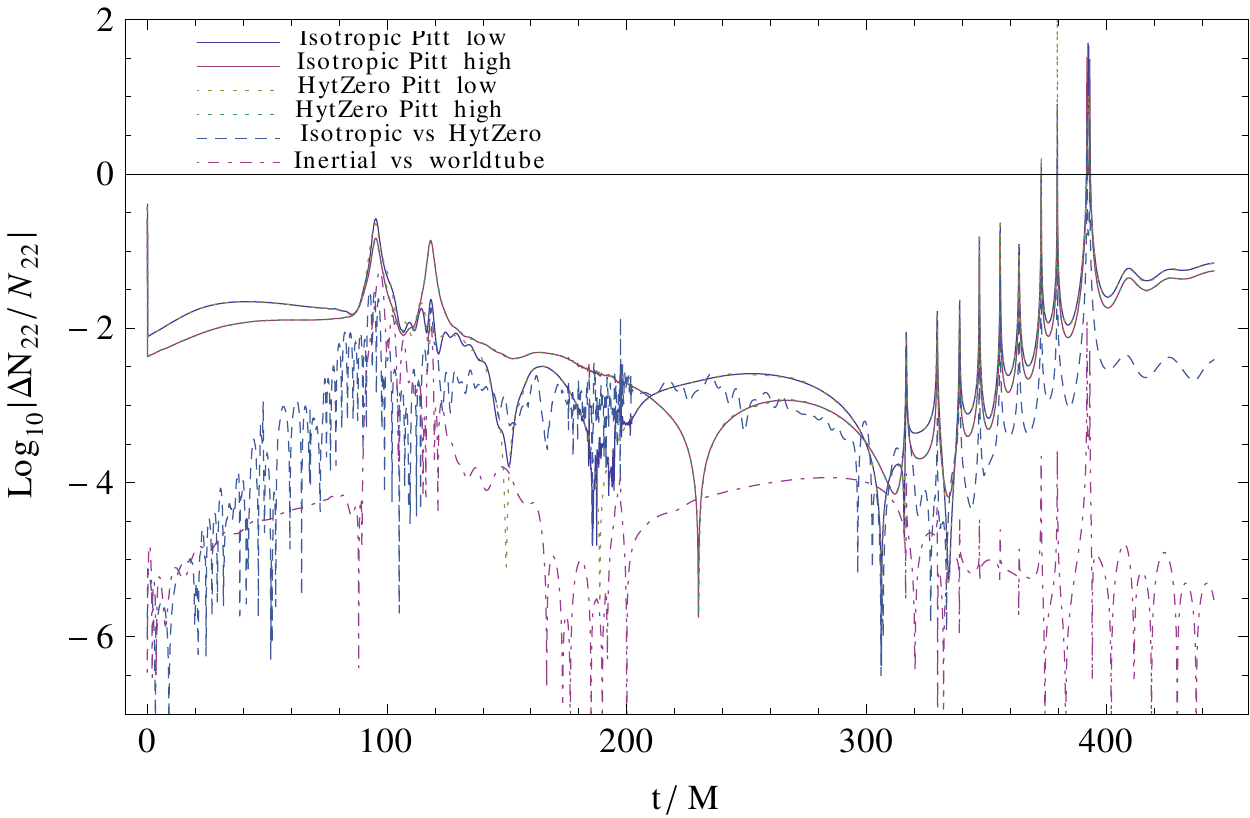}
  \caption{\small{Graphs of the relative error \(\log_{10}|\Delta
      N_{22}/N_{22}|\) in the news
      function for Pitt extraction at \(R=100M\). Again, the errors
      for the Pitt1 and Pitt2 resolutions demonstrate convergence.
      Compared to Fig.~\ref{fig:AGPittConvNews30}, the relative error
      (Inertial vs worldtube) between inertial and worldtube coordinates has now
      dropped below the Isotropic vs ³ÉHytZero gauge effect. }}
      
  \label{fig:AGPittConvNews100}
\end{figure}

\begin{figure}[h!]
  \centering
  \includegraphics[width=0.8\textwidth]{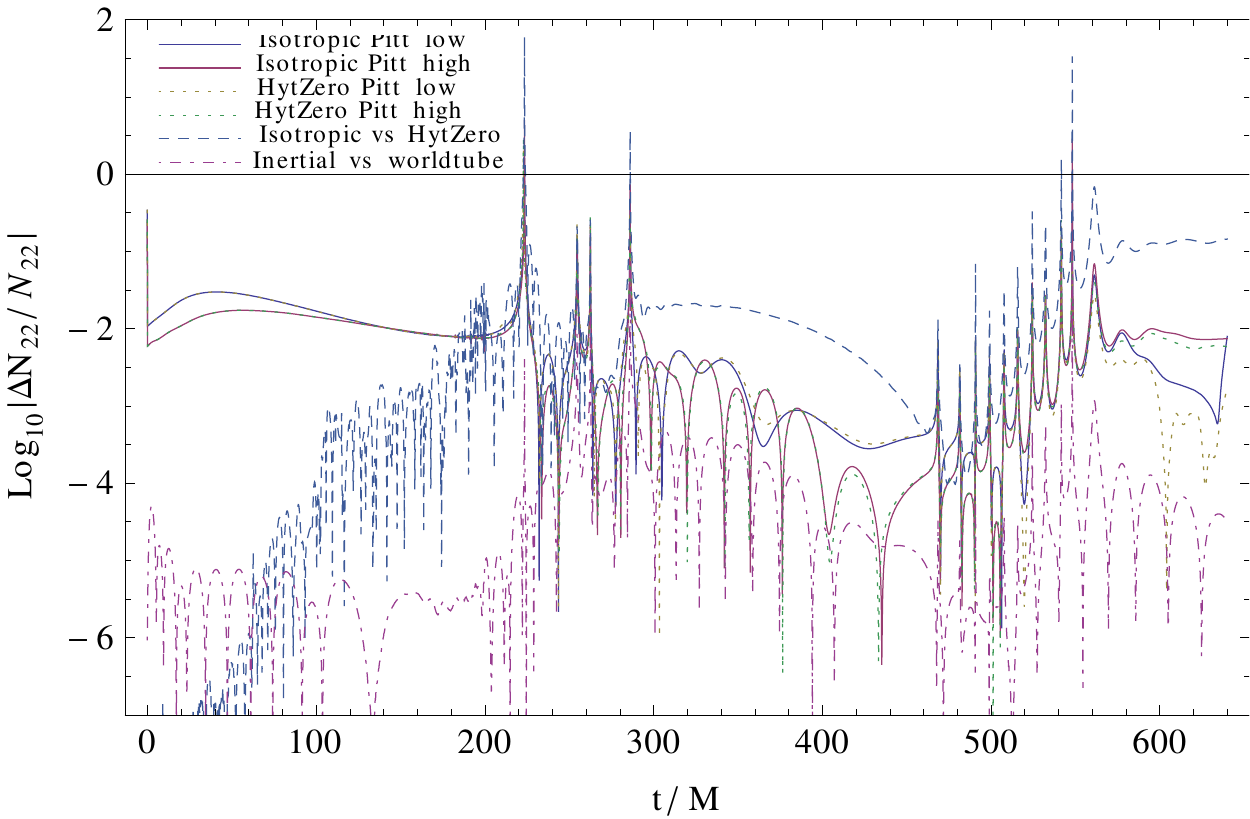}
  \caption{\small{Graphs of the relative error \(\log_{10}|\Delta
      N_{22}/N_{22}|\) in the news
      function for Pitt extraction at \(R=250M\). At this large
      extraction radius, the dominant error arises from the
      Isotropic-HytZero gauge effect. }}
  \label{fig:AGPittConvNews250}
\end{figure}

Here, in order to establish a baseline, we examine the self
convergence of the Pitt code for each of the extraction radii, using
the three resolutions (Pitt1,Pitt2,Pitt3) indicated in
Table~\ref{tab:ALLTHETHINGS}. In Figs.~\ref{fig:AGPittConvNews30},
\ref{fig:AGPittConvNews100}, and \ref{fig:AGPittConvNews250}, we see in
both the Isotropic and HytZero gauges that the news function converges
over the entire run. Indeed, Isotropic (solid lines) and HytZero
(dashed lines) overlap completely. The figures also plot the relative
error in the news computed in both gauges, which is consistently below the
numerical error implied by convergence tests
for extraction worldtubes at \(R=30M\) and \(R=100M\).
This verifies that the Pitt code successfully removes gauge
effects. Furthermore, the figures plot the relative error between the
news computed in the worldtube coordinates and the inertial
coordinates at \(\scri^+\). In~the \(R=30M\) case shown in
Fig.~\ref{fig:AGPittConvNews30}, the initial discrepancy is high due
to the strong gauge effects of junk radiation. It does not fall below
the relative error between the Isotropic and HytZero gauges until well
after the signal has passed. This confirms that the transformation to
inertial coordinates is essential for correctly removing gauge effects
from the waveform. For extraction at \(R=100M\) shown in
Fig.~\ref{fig:AGPittConvNews100}, the relative error between worldtube
and inertial coordinates has dropped below the Isotropic-HytZero gauge
effect. At \(R=250M\) shown in Fig.~\ref{fig:AGPittConvNews250}, the
predominant error is the Isotropic-HytZero gauge effect.

These results show that the selected runs do produce a substantial
gauge error between the worldtube and inertial coordinates and that
the Pitt code effectively removes it, while remaining convergent for
the duration of the run.

\subsection{\texttt{SpEC} code convergence and removal of gauge effects}
\label{sec:spec}

\begin{figure}[h!]
  \centering
  \includegraphics[width=0.8\textwidth]{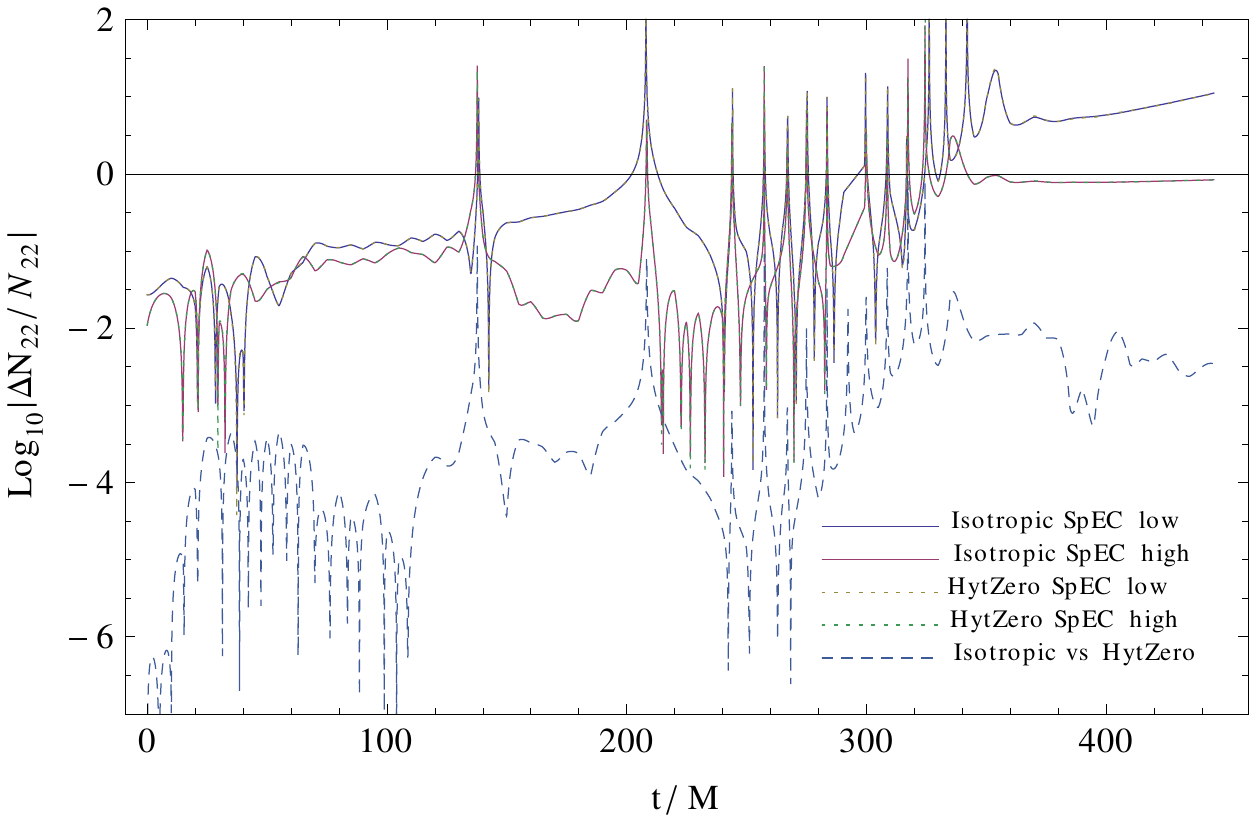}
  \caption{\small{Graphs of the relative error \(\log_{10}|\Delta
      N_{22}/N_{22}|\) in the news function for the \(R=30M\)
      \texttt{SpEC} extraction run in both gauges. The relative errors
      for the \texttt{SpEC}1 (low) and \texttt{SpEC}2 (high) resolutions (compared
      to \texttt{SpEC}2 and \texttt{SpEC}3 respectively) are rescaled
      to demonstrate convergence. The graph Isotropic vs HytZero
      indicates the relative error between the news computed in both
      gauges. Even at this small extraction radius, there is
      relatively little Isotropic vs HytZero gauge error.}}
  \label{fig:AGSpECConvNews30}
\end{figure}

\begin{figure}[h!]
  \centering
  \includegraphics[width=0.8\textwidth]{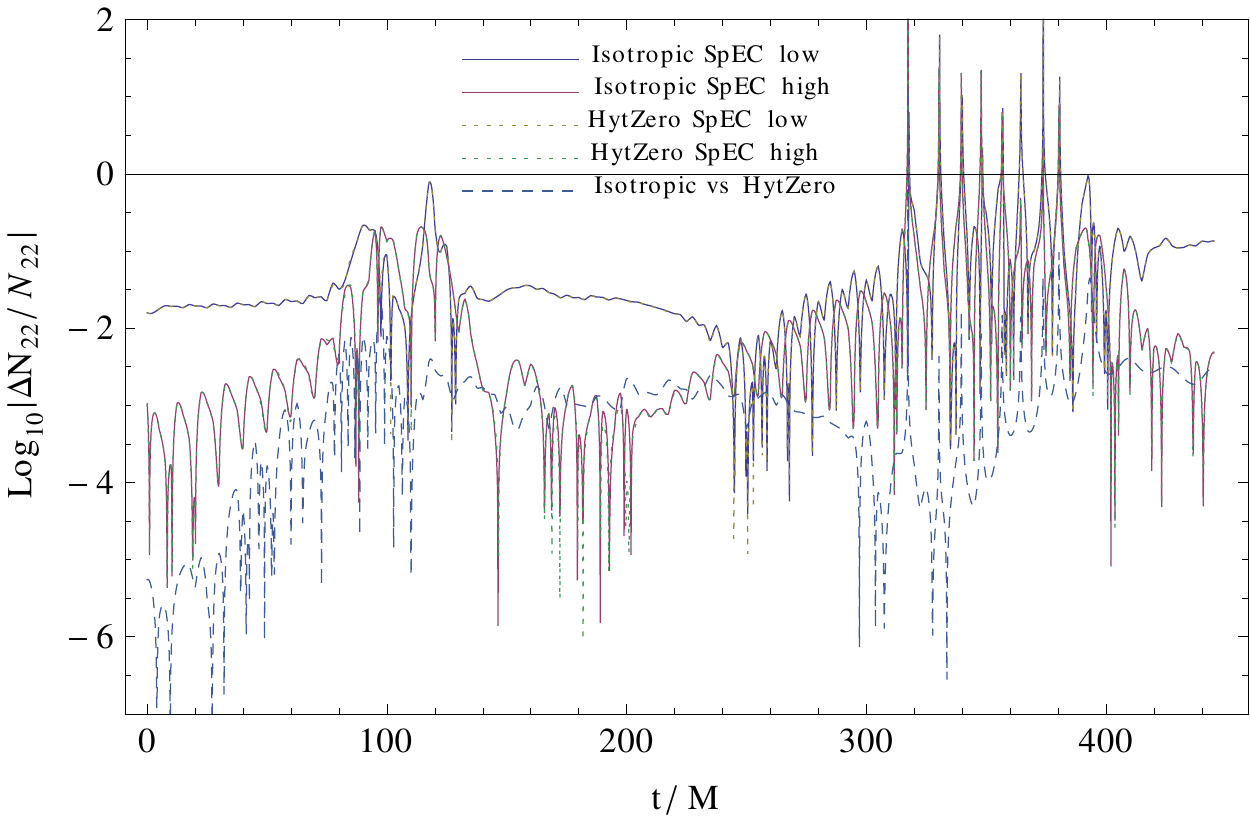}
  \caption{\small{Graphs of the relative error \(\log_{10}|\Delta
      N_{22}/N_{22}|\) in the news function for the \(R=100M\)
      \texttt{SpEC} extraction run in both gauges. The graphs show
      convergence in both gauges. The Isotropic vs HytZero gauge error is
      relatively small.}}
  \label{fig:AGSpECConvNews100}
\end{figure}

\begin{figure}[h!]
  \centering
  \includegraphics[width=0.8\textwidth]{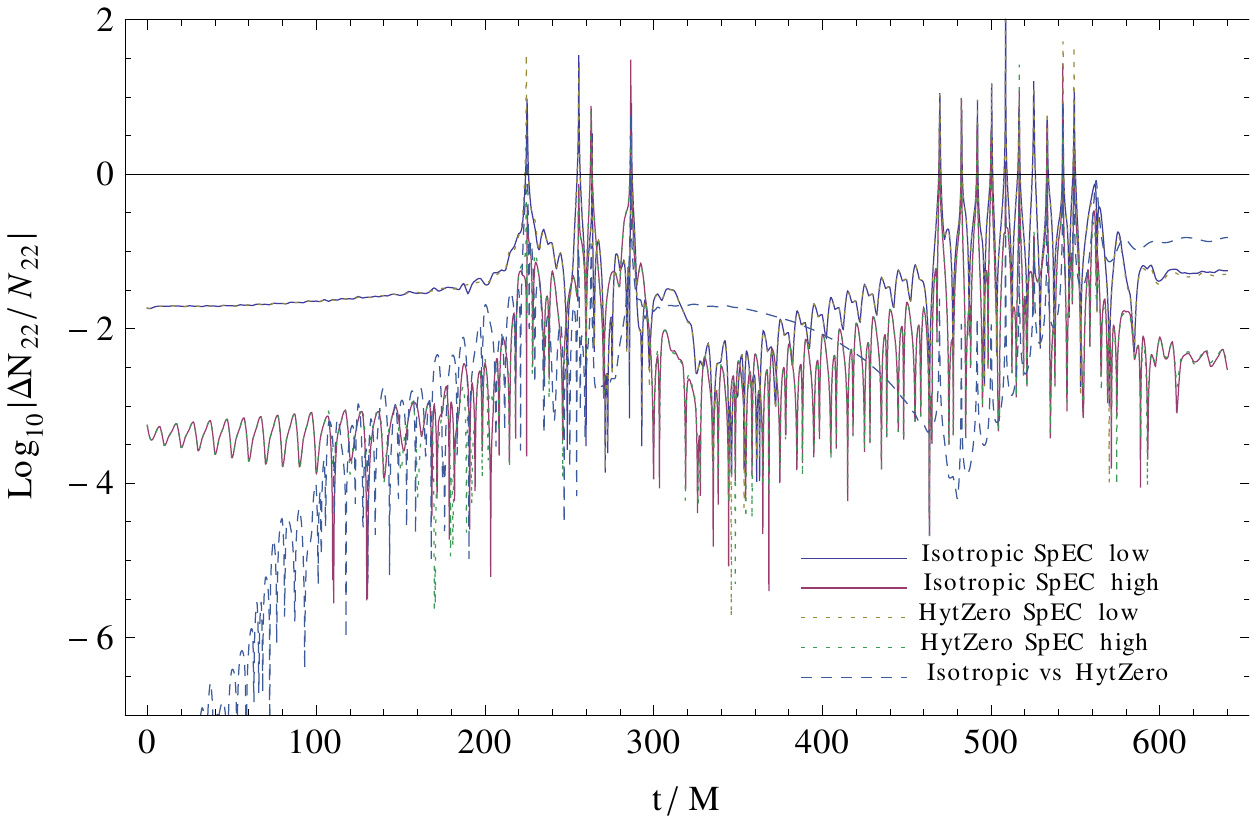}
  \caption{\small{Graphs of the relative error \(\log_{10}|\Delta
      N_{22}/N_{22}|\) in the news
      function for the \(R=250M\) \texttt{SpEC} extraction run in both
      gauges, showing convergence in both gauges as well as small
      gauge error throughout the run.}}
  \label{fig:AGSpECConvNews250}
\end{figure}

Here we examine the \texttt{SpEC} code's self convergence for each
extraction radii, in the same way that the Pitt code was examined in
Sec.~\ref{sec:pitt}. In Figs.~\ref{fig:AGSpECConvNews30},
\ref{fig:AGSpECConvNews100}, and \ref{fig:AGSpECConvNews250}, we see
that convergence, measured with the resolutions indicated in
Table~\ref{tab:ALLTHETHINGS}, is comparable to the Pitt code's
convergence, while the potential gauge contamination is consistently
removed at all worldtube radii. As in
Figs.~\ref{fig:AGPittConvNews30}, \ref{fig:AGPittConvNews100}, and
\ref{fig:AGPittConvNews250}, the solid lines (Isotropic) and dashed
lines (HytZero) overlap due to consistency in gauge removal. The
\texttt{SpEC} extraction code
effectively removes gauge error at all radii while remaining
convergent throughout the runs.

\subsection{Comparison of \(\Psi_4^0\) between
the Pitt and \texttt{SpEC} codes}

In Secs~\ref{sec:pitt} and \ref{sec:spec}, we have shown that both
codes are convergent and remove potential gauge effects. We have also
demonstrated that the difference in the news computed by the two codes
disappears as the extraction worldtube radius increases. Here we
provide further evidence that even at a small worldtube radius the
waveform computed by the \texttt{SpEC} code is valid.

After the gauge freedom is removed by extraction, there is still
supertranslation and Lorentz freedom in the choice of inertial
coordinates, which affect the phase and velocity of the inertial
observers. This effect is highly sensitive to initial conditions and
also to the evolution of the inertial confomal transformation factor
\(\omega\), especially in the extreme gauge conditions of extraction
at \(R=30M\). It feeds into the worldtube interpolation error in the
Pitt code. In order to verify that the discrepancy illustrated in
Fig.~\ref{fig:AGNewsWaveform30} between the news computed by the Pitt
null code and \texttt{SpEC} is partially due to this inertial
coordinate freedom, we compute the time derivative of the news, which
is related to the Weyl curvature in inertial coordinates according to
\(\partial_t N = \Psi_4^0\). This suppresses phase differences between
the two waveforms. In making the comparisons, \(\Psi_4^0\) is computed
semi-independently using the Weyl tensor waveform module in the current
version of the Pitt code~\cite{Babiuc:2010ze}. In these runs,
\(\Psi_4^0\) was found to be convergent with truncation error
comparable to the consistency between \(\Psi_4^0\) and \(\partial_t
N\) in the Pitt code.

\begin{figure}[h!]
    \centering
    \includegraphics[width=0.8\textwidth]{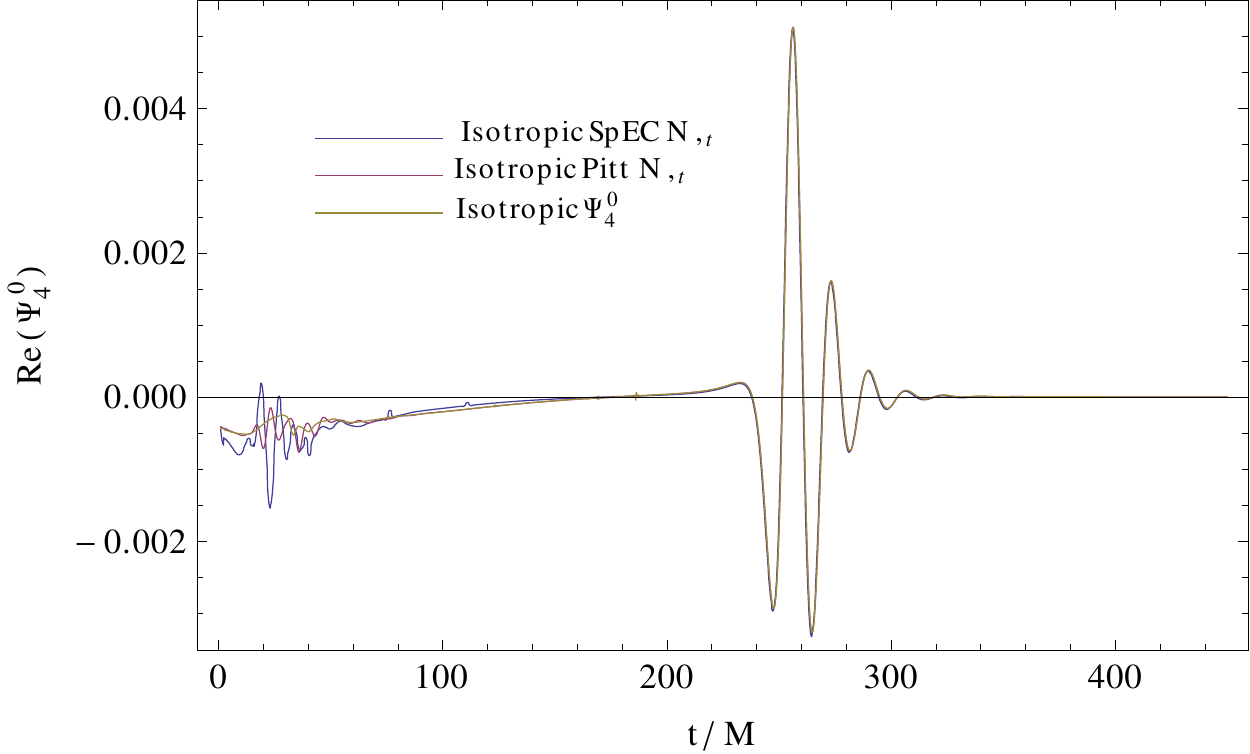}
    \caption{\small{The real part of the \((2,2)\) modes of the
        waveforms of \(\Psi_4^0\) (computed during the Pitt run) and the
        time derivative of the news \(\partial_t N \), as computed for
        the \texttt{SpEC} and Pitt runs using the Isotropic gauge with
        extraction radius \(R=30M\). The waveforms are in much better
        agreement than the comparison of the news waveforms for this
        extraction radius in Fig.~\ref{fig:AGNewsWaveform30}. }}
    \label{fig:AGPsi4Waveform30}
\end{figure}

\begin{figure}[h!]
    \centering
    \includegraphics[width=0.8\textwidth]{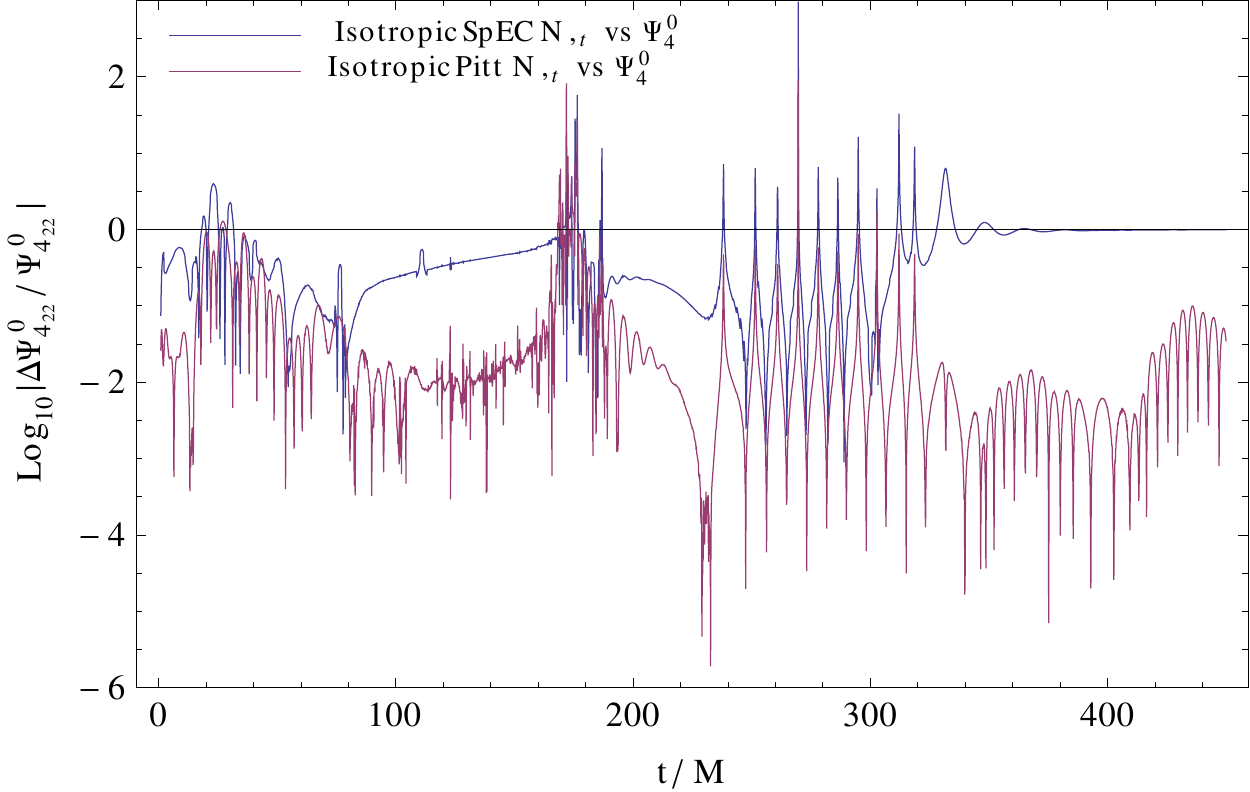}
    \caption{\small{The relative error \(\log_{10}|\Delta
        \Psi^0_{4_{22}}/\Psi^0_{4_{22}}|\) between \(\Psi_4^0\) and
        \(\partial_t N\) as computed for the \texttt{SpEC} and Pitt runs
        using the Isotropic gauge with extraction radius \(R=30M\),
        corresponding to the waveforms in
        Fig.~\ref{fig:AGPsi4Waveform30}. \(\Psi_4^0\) and its numerical
        truncation error were computed using the Pitt code. Truncation
        error in \(\Psi_4^0\) was consistent with the \(\partial_t N\)
        error shown here.  The good agreement between the \texttt{SpEC}
        and Pitt results eventually lapses, but not until well after
        ring-down. }}
    \label{fig:AGConvPsi430}
\end{figure}

\begin{figure}[h!]
    \centering
    \includegraphics[width=0.8\textwidth]{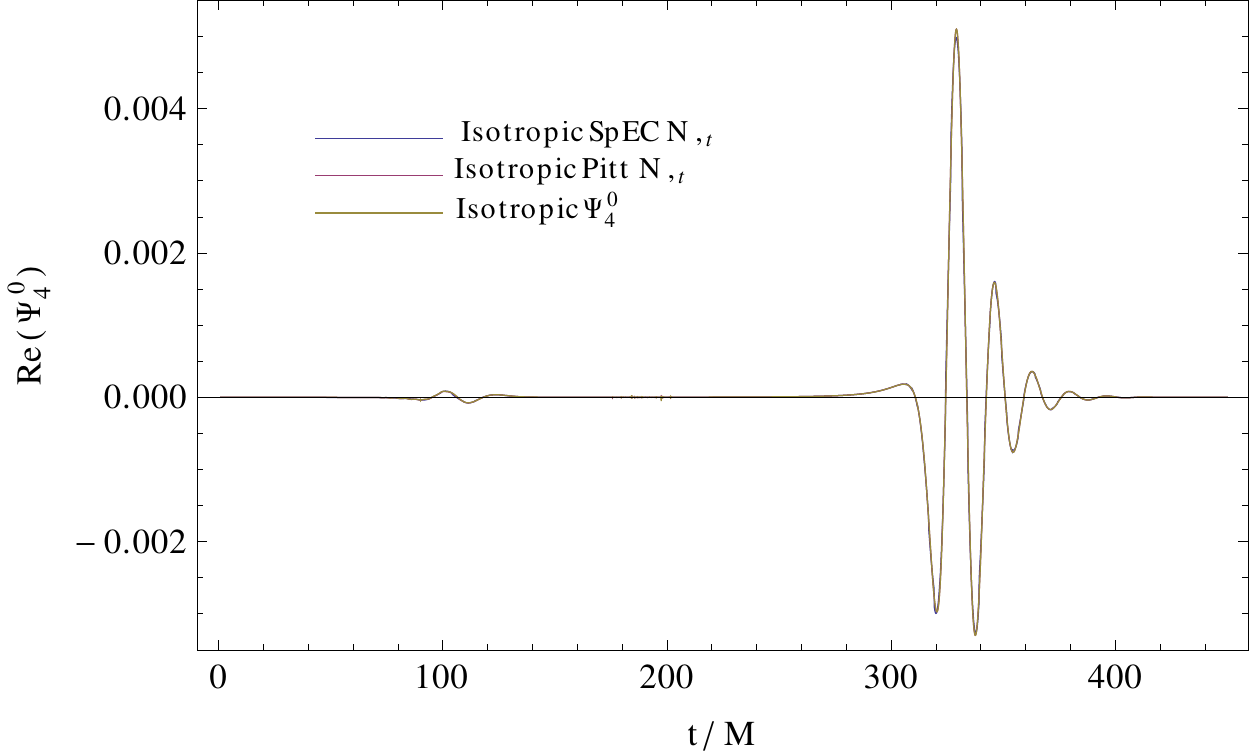}
    \caption{\small{Waveforms of \(\Psi_4^0\) and \( \partial_t N \) as
        computed for the \texttt{SpEC} and Pitt runs using the Isotropic
        gauge with extraction radius \(R=100M\).}}
    \label{fig:AGPsi4Waveform100}
\end{figure}

\begin{figure}[h!]
    \centering 
    \includegraphics[width=0.8\textwidth]{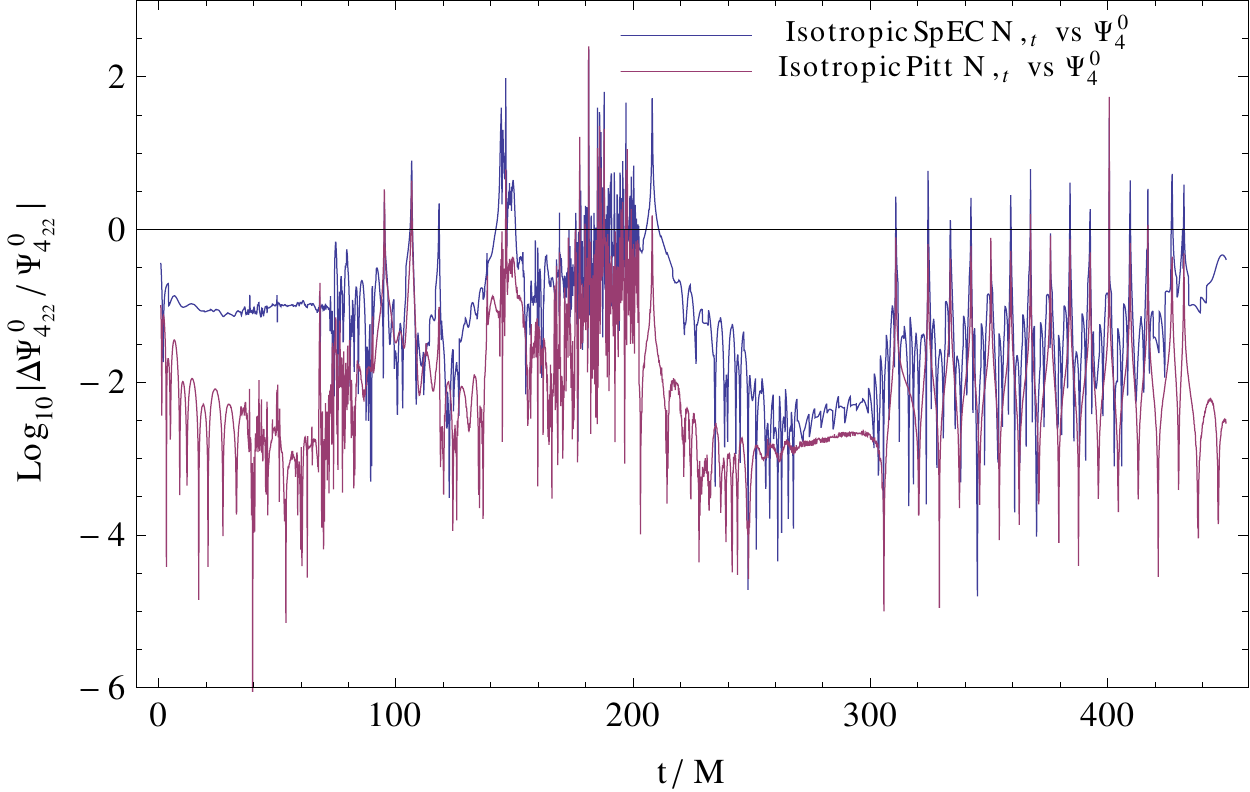}
    \caption{\small{The relative error \(\log_{10}|\Delta
        \Psi^0_{4_{22}}/\Psi^0_{4_{22}}|\) between \(\Psi_4^0\)
        and \(\partial_t N\) as computed for the \texttt{SpEC} and Pitt
        runs using the Isotropic gauge with extraction radius
        \(R=100M\), corresponding to the
        waveforms in Fig.~\ref{fig:AGPsi4Waveform100}.
        Both codes show comparable agreement throughout the
        run.}}
    \label{fig:AGConvPsi4100}
\end{figure}

\begin{figure}[h!]
    \centering 
    \includegraphics[width=0.8\textwidth]{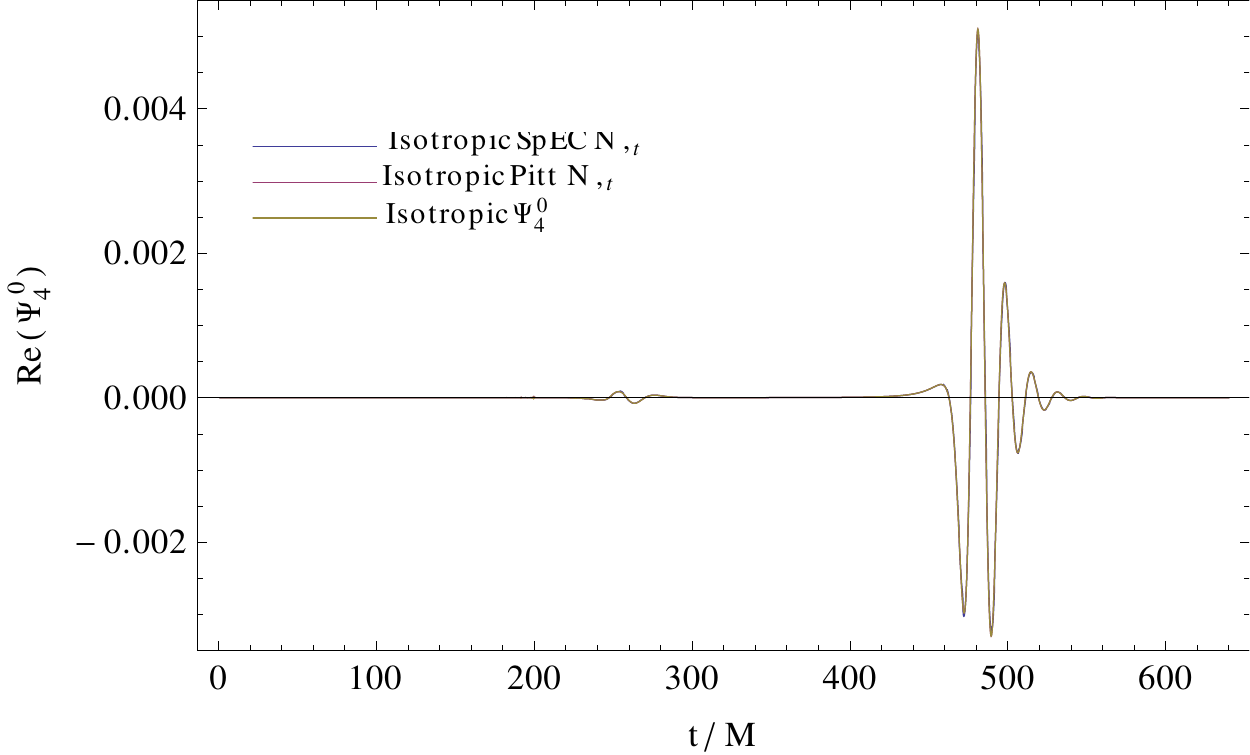}
    \caption{\small{Waveforms of \(\Psi_4^0\) and \(\partial_t N\) as
        computed for the \texttt{SpEC} and Pitt codes using the
        Isotropic gauge with extraction radius \(R=250M\).}}    
    \label{fig:AGPsi4Waveform250}
\end{figure}

\begin{figure}[h!]
    \centering 
    \includegraphics[width=0.8\textwidth]{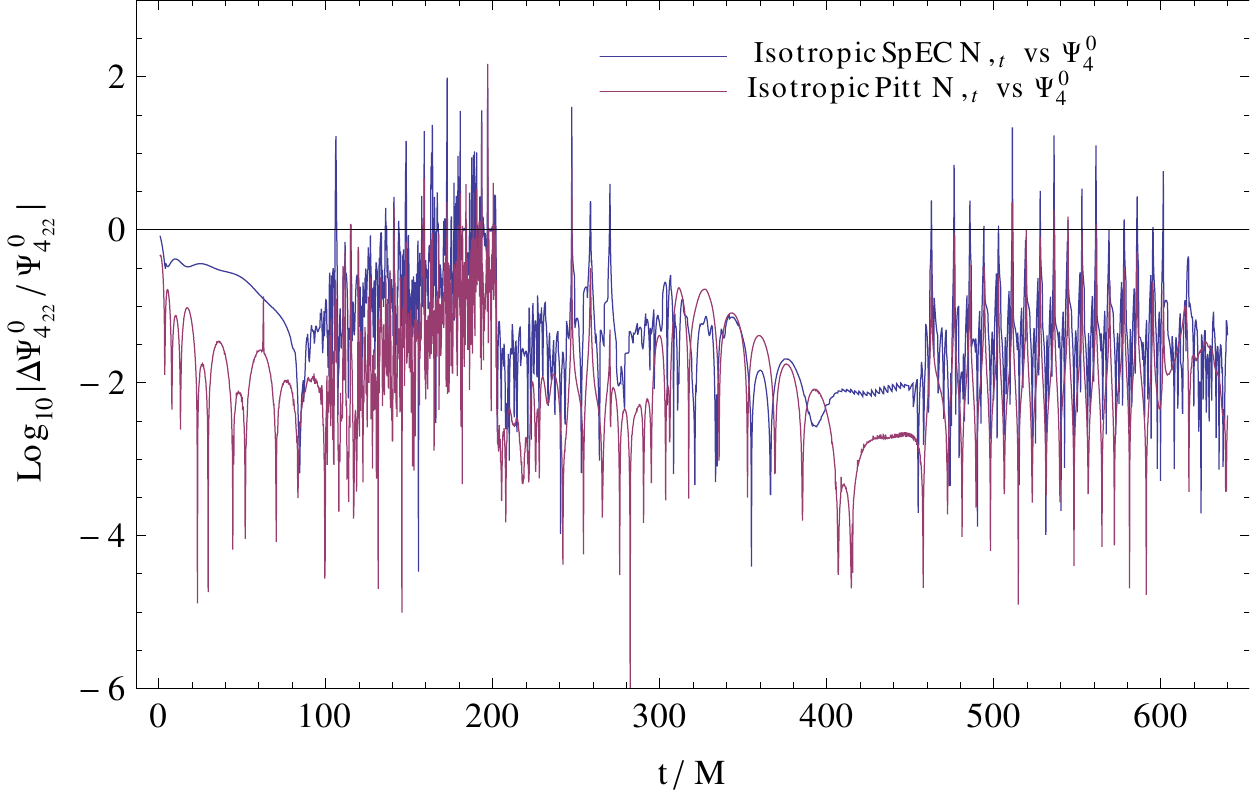}
    \caption{\small{The relative error \(\log_{10}|\Delta
        \Psi^0_{4_{22}}/\Psi^0_{4_{22}}|\) between \(\Psi_4^0\) and
        \(\partial_t N\) computed for the \texttt{SpEC} and Pitt runs
        using the Isotropic gauge with extraction radius \(R=250M\),
        corresponding to the waveforms in
        Fig.~\ref{fig:AGPsi4Waveform250}. Both codes show good
        agreement throughout the run.}}
    \label{fig:AGConvPsi4250}
\end{figure}

In Fig.~\ref{fig:AGPsi4Waveform30}, we see that the time derivative of
the news and \(\Psi_4^0\) have much less discrepancy than
Fig.~\ref{fig:AGNewsWaveform30} would suggest. In
Figs.~\ref{fig:AGConvPsi430}, \ref{fig:AGConvPsi4100}, and
\ref{fig:AGConvPsi4250}, we compare relative errors between
\(\Psi_4^0\) and \(\partial_t N\) computed by the Pitt and \texttt{SpEC}
codes. Not only is there agreement between the codes at \(R=30M\),
this agreement persists for larger extraction radii, as shown in
Figs.~\ref{fig:AGPsi4Waveform100} and \ref{fig:AGPsi4Waveform250}.
Both codes show agreement with the \(\Psi_4^0\) waveform throughout
the runs at all three extraction radii. This indicates that a major part of the
discrepancy in Fig.~\ref{fig:AGNewsWaveform30} is due to initialization
errors in the Pitt code, confirming the physical validity of
the extracted \texttt{SpEC} waveform.




\subsection{Relative motion between inertial and worldtube coordinates}

\begin{figure}[h!]
    \centering 
    \includegraphics[width=0.8\textwidth]{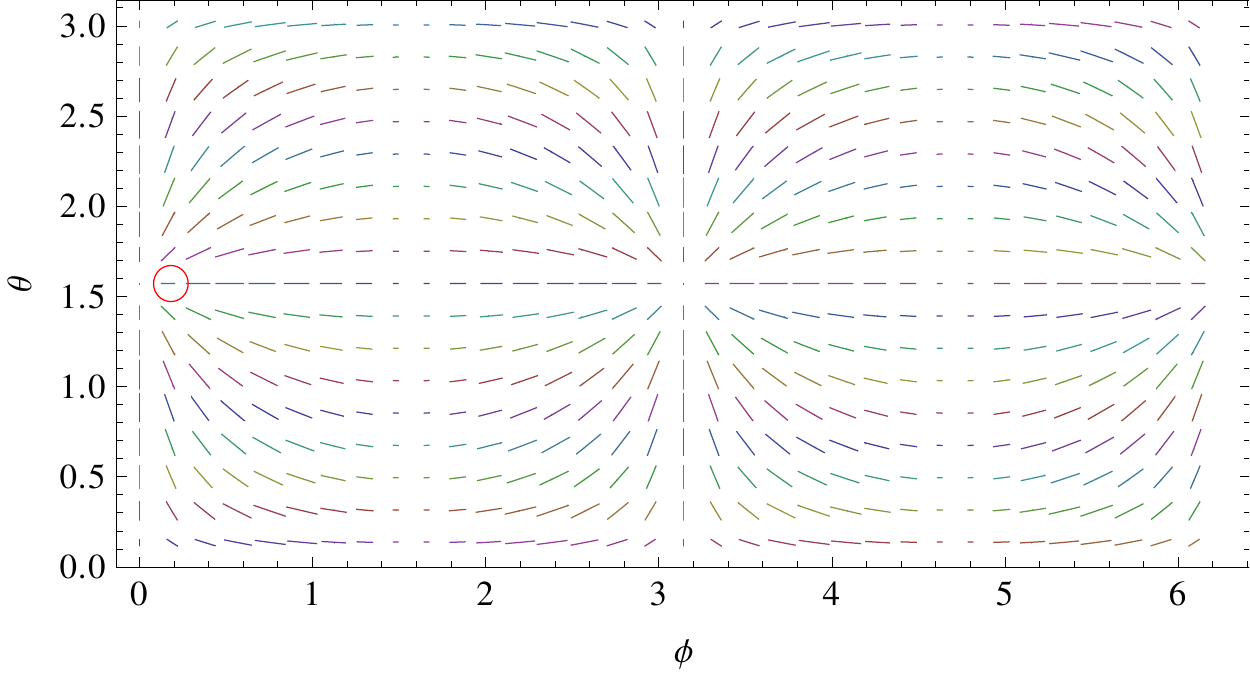}
    \caption{\small{The arrows indicate the global pattern of the motion
        of the \((\theta,\phi)\) inertial coordinates at each spherical
        collocation point (labeled in radians) relative to the worldtube
        angular coordinates. Global motion is primarily in the Cartesian
        \(x\) direction, and has been exaggerated by a factor of 10 for
        visibility.}}     
    \label{fig:AGInerCoords}
\end{figure}

In Section \ref{sec:waveforms}, we discussed the construction of an
inertial coordinate system and its evolution with respect to the
worldtube coordinates constructed from the Cartesian coordinates of
the Cauchy code. Here, we describe the motion of the inertial \((\theta,\phi)\)
angular coordinates relative to the worldtube angular coordinates,
constructed in the standard way from the worldtube Cartesian coordinates.
Figure \ref{fig:AGInerCoords} illustrates the global pattern of this relative
motion for the Isotropic gauge \texttt{SpEC} run with the highest
resolution extraction at \(R=30M\). Generally speaking, the
coordinates wiggle back and forth in the direction corresponding to
the motion of the black holes. The complete movement amounts to at
most a few percent of their initial values, but even this is
sufficient to introduce considerable gauge error in the waveform, as
already seen in Fig.~\ref{fig:AGPittConvNews30}.

\begin{figure}[h!]
    \centering
    \includegraphics[width=0.8\textwidth]{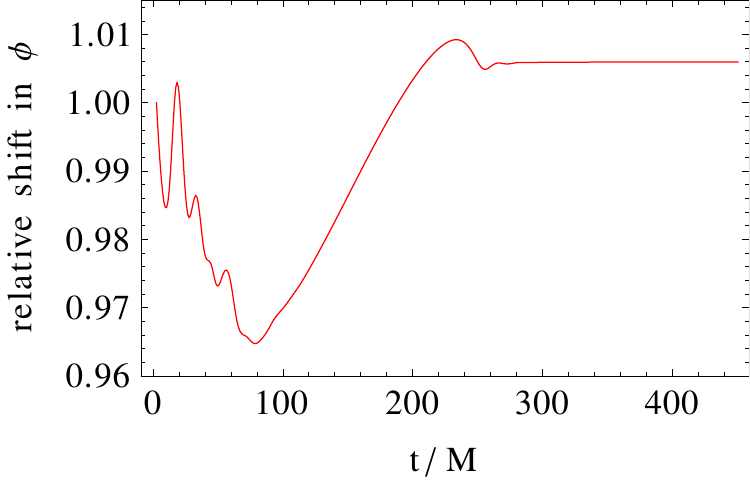}
    \caption{\small{ The relative inertial-worldtube \(\phi\)-coordinate
        motion of the point circled in
        Fig.~\ref{fig:AGInerCoords} shows approximately a \(3.5\)  percent
        variation from its initial value.}}
    \label{fig:AGInerCoordvsTime}
\end{figure}

In Fig.~\ref{fig:AGInerCoordvsTime}, the relative \(\phi\) motion of
the point circled in Fig.~\ref{fig:AGInerCoords} is plotted as a
function of inertial time. Initial junk radiation causes considerable
wobble, followed by a smooth return almost to its starting point. The
maximum excursion of the \(\phi\)-coordinate shift from
its initial value is about 3.5\%.

\subsection{Precessing, spinning binary black hole merger}

\begin{figure}[h!]
  \centering
  \includegraphics[width=0.8\textwidth]{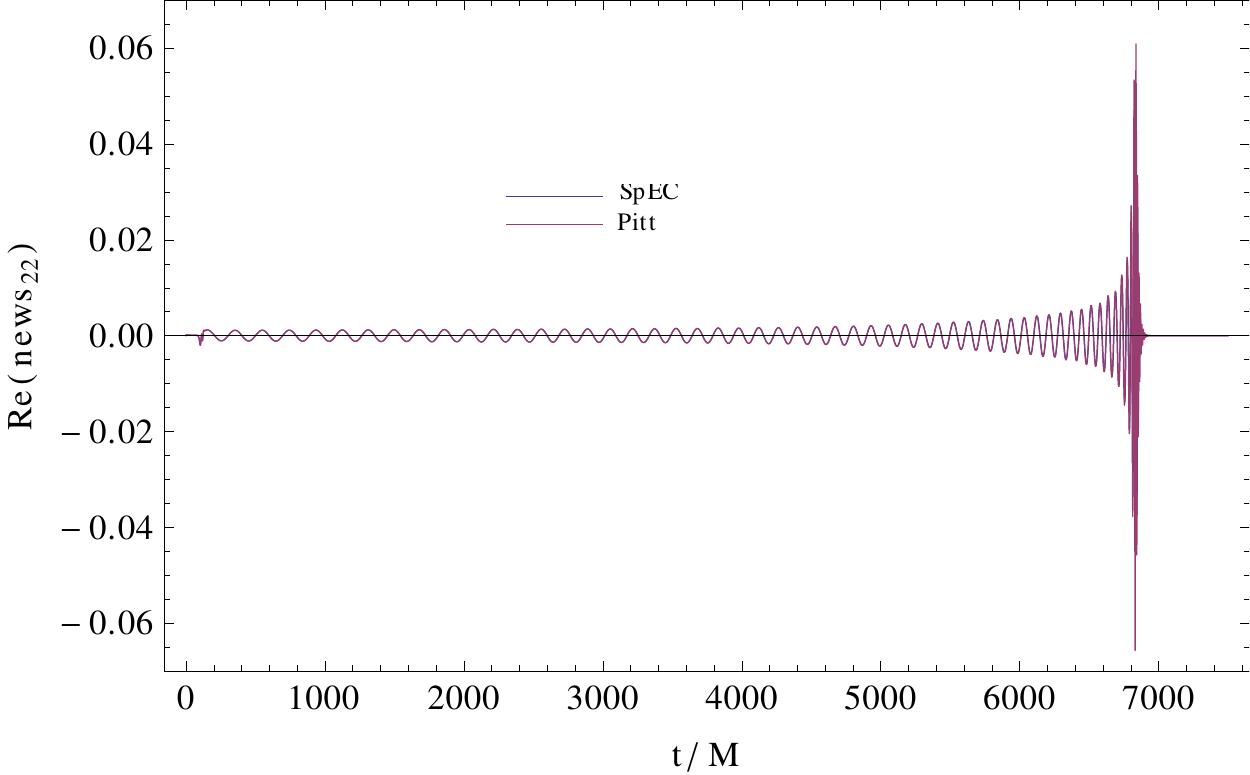}
  \caption{\small{Waveform of the real part of the \((2,2)\) mode of the
      news function for the generic precessing binary black hole run,
      showing 26 orbits, inspiral, merger, and ringdown.}}
  \label{fig:AGGenericNewsWaveform}
\end{figure}

In addition to the head-on collision tests which we have described, we
have also investigated stability and convergence of the Pitt and
\texttt{SpEC} CCE modules, together with the
inertial-worldtube coordinate transformation, using the generic test
run of precessing, spinning binary black holes in \cite{Handmer:2014},
as taken from Taylor {\it et al.}\cite{Taylor:2013zia}. Its parameters
are mass ratio \(q=3\), black hole spins \(\chi_1 =
(0.7,0,0.7)/\sqrt{2}\) and \(\chi_2 = (-0.3,0,0.3)/\sqrt{2}\), number
of orbits 26, total time \(T=7509M\), initial eccentricity
\(10^{-3}\), initial frequency \(\omega_{ini} = 0.032/M\), and
extraction radius \(R=100M\). The Pitt and \texttt{SpEC} waveforms 
displayed in Fig.~\ref{fig:AGGenericNewsWaveform}
are fairly typical waveforms, spanning the initial junk radiation through
inspiral, merger, and ringdown.

\begin{figure}[h!]
    \centering
    \includegraphics[width=0.8\textwidth]{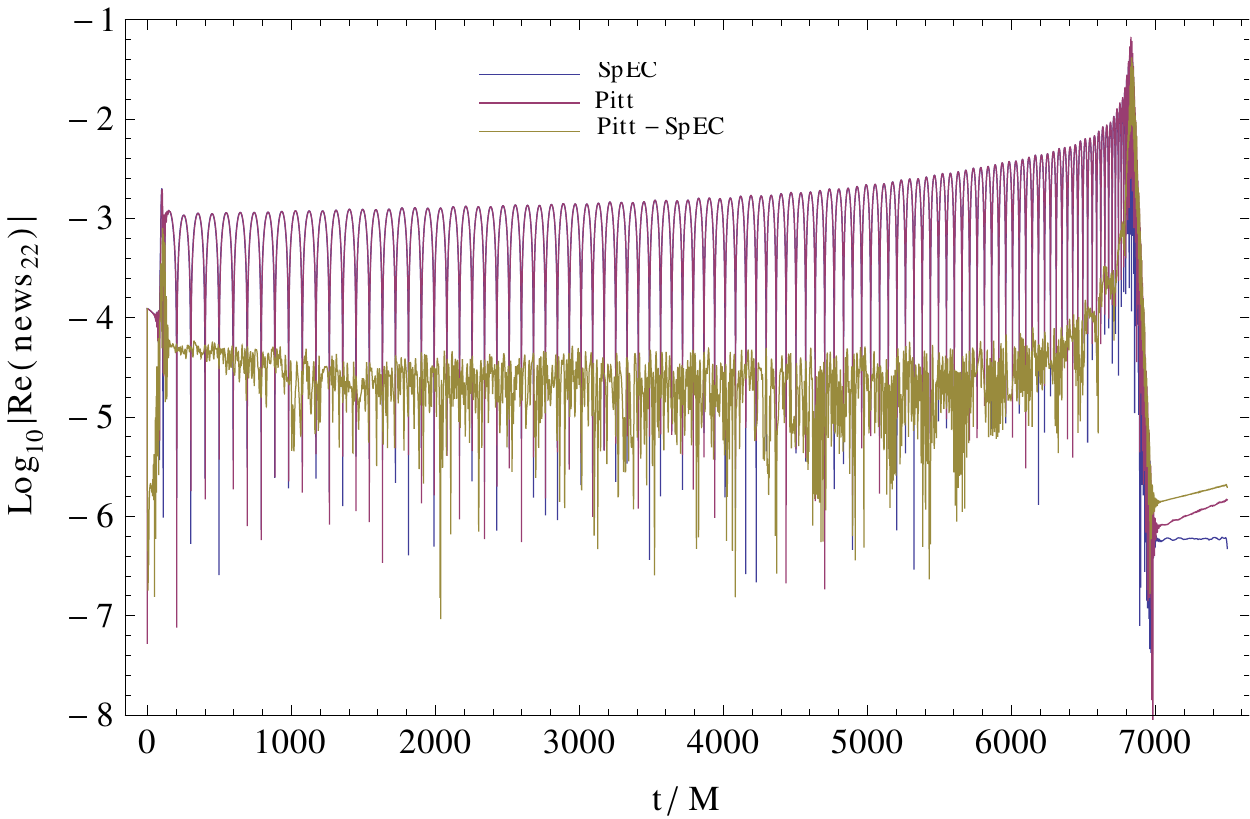}
    \caption{\small{Graphs of \(\log_{10}|Re(\text{news}_{22})|\) for
        Pitt and \texttt{SpEC} waveforms for the precessing binary
        black hole run, showing 16 orbits, inspiral, merger, and
        ringdown. The difference \(\textrm{Pitt}-\textrm{SpEC}\) gives
        the absolute error \(\log_{10}|N_{22_{Pitt}} -
        N_{22_{SpEC}}|\) between the codes, showing consistent
        agreement throughout the run.}}
    \label{fig:AGGenericNewsWaveformLog}
\end{figure}

\begin{figure}[h!]
    \centering
    \includegraphics[width=0.8\textwidth]{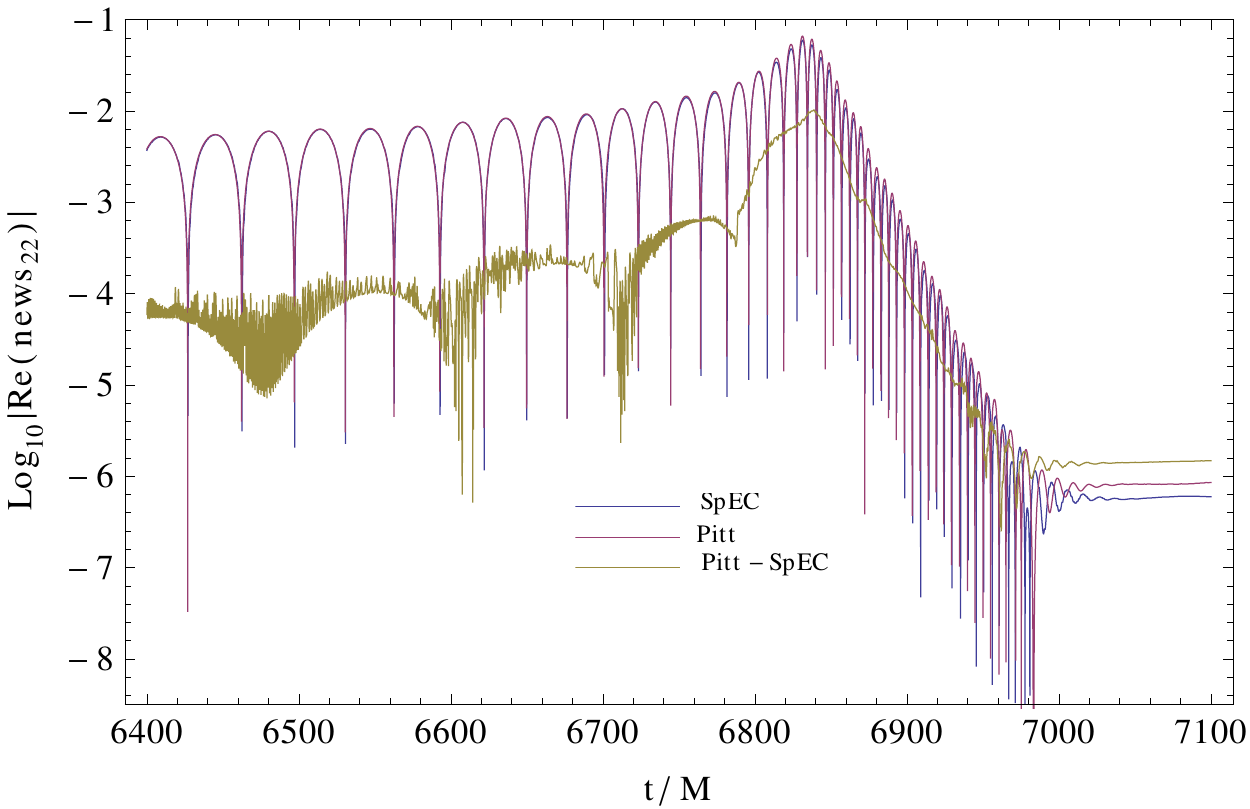}
    \caption{\small{Same graphs as in
        Fig.~\ref{fig:AGGenericNewsWaveformLog} but focusing on the merger
        part of the waveform. Absolute error shows consistency between the Pitt
        and \textrm{SpEC} news waveforms
        through merger and ringdown.}}
    \label{fig:AGGenericNewsWaveformLogMerger}
\end{figure}

Figures~\ref{fig:AGGenericNewsWaveformLog} and
\ref{fig:AGGenericNewsWaveformLogMerger} show a log scale comparison
of the waveforms with absolute error. 
The codes agree strongly throughout the run.


\begin{figure}[h!]
  \centering
  \includegraphics[width=0.8\textwidth]{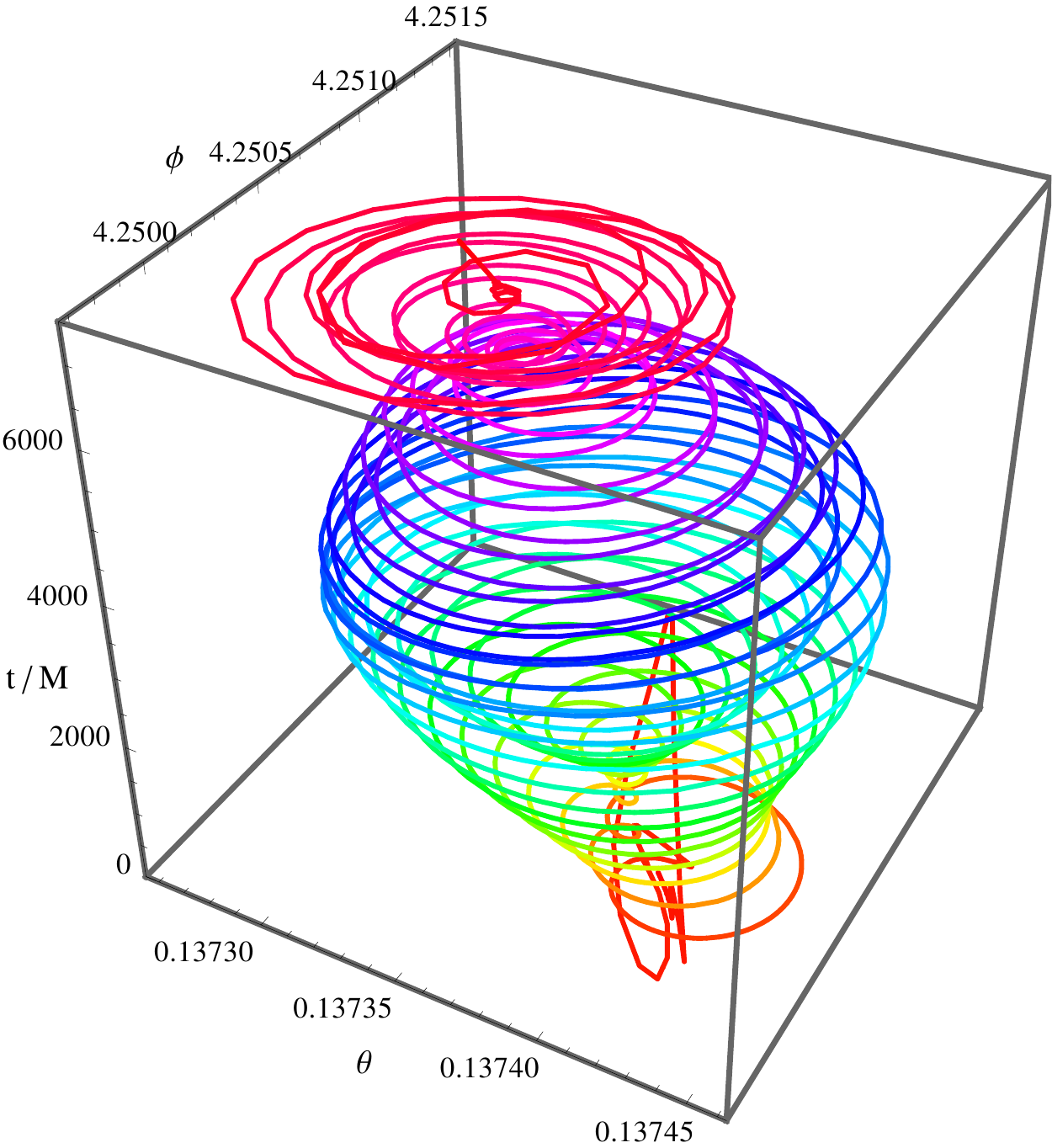}
  \caption{\small{Relative motion of a representative angular coordinate taken from \texttt{SpEC}3 with extraction radius \(R=100M\)
      showing a long term
      helical pattern concordant with the black hole inspiral and
      merger. Globally, initial oscillation due to junk radiation is
      aligned primarily along the Cartesian \(x\) direction,
      corresponding to the initial black hole orientation. Coordinate
      motion is an epicyclic helix whose amplitude is modulated by
      precession of the orbital plane.}}
  \label{fig:AGGenericInerCoordvsTime}
\end{figure}

The relative inertial-worldtube coordinate motion of a representative
point in the extended generic run is illustrated in
Fig.~\ref{fig:AGGenericInerCoordvsTime}. At early times, the helix has
two loops per cycle corresponding to each of the black holes. At later
times, precession dominates the evolution of this particular
coordinate. Throughout the run, the deviation is around 0.5\%.

\section{Conclusion}

The \texttt{SpEC} characteristic evolution algorithm has now been
furnished with a convergent, efficient news extraction
module. \texttt{SpEC} is now capable of rapidly producing accurate,
gauge free waveforms as required.

\ack{We thank Nicholas Taylor for his generic spinning binary
  black hole run that we used to
  test and baseline code performance. We thank Yosef Zlochower for
  supplying details of the news module in the Pitt null code. We thank
  Mark Scheel, Yanbei Chen, and Christian Reisswig for their advice,
  support, and technical expertise,
  and thank Michael Boyle for comments on the manuscript.
  This research used the Spectral
  Einstein Code (\texttt{SpEC})\cite{Mroue:2013PRL}. The Caltech
  cluster \texttt{zwicky.cacr.caltech.edu} is an essential resource
  for \texttt{\texttt{SpEC}} related research, supported by the
  Sherman Fairchild Foundation and by NSF award PHY-0960291. This
  research also used the Extreme Science and Engineering Discovery
  Environment (XSEDE) under grant TG-PHY990002. The UCSD cluster
  \texttt{ccom-boom.ucsd.edu} was used during code development. This
  project was supported by the Sherman Fairchild Foundation, and by
  NSF Grants PHY-1068881, AST-1333520, and CAREER Grant PHY-0956189 at
  Caltech. JW's research was supported by NSF grant PHY-1201276 to
  the University of Pittsburgh.

\section*{References}

\bibliographystyle{utphys} \bibliography{References/References}

\end{document}